\documentclass[twocolumn,a4paper]{article}
\usepackage{array}
\usepackage{bm}
\usepackage{amsmath}
\usepackage{amssymb}
\usepackage{graphicx}
\usepackage{fancyhdr}
\usepackage[unicode=true,pdfusetitle,
 bookmarks=true,bookmarksnumbered=false,bookmarksopen=false,
 breaklinks=false,pdfborder={0 0 1},backref=false,colorlinks=false]
 {hyperref}
\hypersetup{
 unicode=true,bookmarks=true,bookmarksnumbered=true, bookmarksopen=true,bookmarksopenlevel=2, breaklinks=false, colorlinks=true,linkcolor=magenta, citecolor=blue, urlcolor=blue, filecolor=blue}

\makeatletter

\providecommand{\tabularnewline}{\\}

\newcommand{\lyxaddress}[1]{
\par {\raggedright #1
\vspace{1.4em}
\noindent\par}
}

\usepackage[usedoi,linkdoi]{rsc}
\usepackage[scale={0.8,0.8}]{geometry}
\usepackage{amsmath,bm}
\usepackage{xcolor}

\makeatother

\begin{document}
\global\long\def\vr{\bm{r}}
\global\long\def\vR{\bm{R}}
\global\long\def\vk{\bm{k}}
\global\long\def\vK{\bm{K}}

\global\long\def\bktwo#1#2{\langle#1|#2\rangle}

\global\long\def\bkthree#1#2#3{\langle#1|#2|#3\rangle}

\global\long\def\ket#1{|#1\rangle}
\global\long\def\bra#1{\langle#1|}

\global\long\def\ave#1{\langle#1\rangle}

\global\long\def\kp{k\cdot p}

\title{Electronic structures and theoretical modelling of two-dimensional
group-VIB transition metal dichalcogenides}

\author{Gui-Bin Liu$^{1,2}$, Di Xiao$^{3}$, Yugui Yao$^{1}$, Xiaodong
Xu$^{4,5}$, Wang Yao$^{2,*}$}

\maketitle

\thispagestyle{fancy} \chead{[Chem. Soc. Rev., 2015, \textbf{44}, 2643--2663  DOI: \href{http://dx.doi.org/10.1039/C4CS00301B}{10.1039/C4CS00301B}] \\ Reproduced by permission of The Royal Society of Chemistry}

\lyxaddress{$^{1}$School of Physics, Beijing Institute of Technology, Beijing
100081, China}

\lyxaddress{$^{2}$Department of Physics and Center of Theoretical and Computational
Physics, The University of Hong Kong, Hong Kong, China}

\lyxaddress{$^{3}$Department of Physics, Carnegie Mellon University, Pittsburgh,
Pennsylvania 15213, USA}

\lyxaddress{$^{4}$Department of Physics, University of Washington, Seattle,
Washington, USA}

\lyxaddress{$^{5}$Department of Material Science and Engineering, University
of Washington, Seattle, Washington, USA}

\lyxaddress{$^{*}$Corresponding author, wangyao@hku.hk}
\begin{abstract}
Atomically thin group-VIB transition metal dichalcogenides (TMDs)
have recently emerged as a new class of two-dimensional (2D) semiconductors
with extraordinary properties including the direct band gap in the
visible frequency range, the pronounced spin-orbit coupling, the ultra-strong
Coulomb interaction, and the rich physics associated with the valley
degree of freedom. These 2D TMDs exhibit great potentials for device
applications and have attracted vast interest for the exploration
of new physics. 2D TMDs have complex electronic structures which underlie
their physical properties. Here we review the bulk electronic structures
in these new 2D materials as well as the theoretical models developed
at different levels, along which we sort out the understandings on
the origins of a variety of properties observed or predicted.
\end{abstract}

\section{Introduction}

\begin{figure*}
\centering{}\includegraphics[width=14cm]{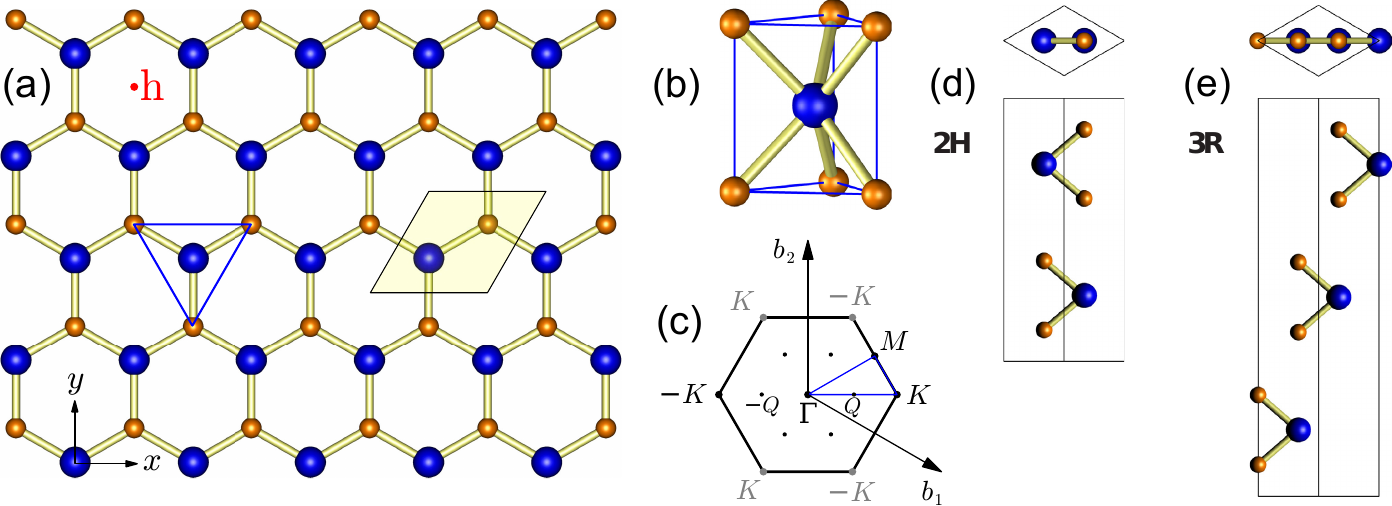}\caption{(a) Top view of group-VIB TMD monolayers. The blue and orange spheres represent
M and X atoms respectively. The light yellow diamond region is the
2D unit cell with lattice constant $a$. (b) Trigonal prismatic coordination
geometry, corresponding to the blue triangle in top view. (c) The
first Brillouin zone. $\bm{b}_{1}$ and $\bm{b}_{2}$ are the reciprocal
lattice vectors. (d) Top and side views of the unit cell of bulk or
bilayer of 2H stacking. (e) Unit cell of bulk group-VIB TMDs of 3R stacking.
\label{fig:struct}}
\end{figure*}

Atomically thin two-dimensional (2D) forms of layered transition metal dichalcogenides (TMDs) have recently attracted remarkable scientific and technological interest.\citep{novoselov_two_dimensional_2005,Zeng_Zhang_2011_50_11093__Single,Coleman_et_2011_331_568__Two,Geim_Grigorieva_2013_499_419__Van} These TMDs have the chemical composition of MX$_{2}$, where M stands for the transition metal elements and X for the chalcogen elements. They exhibit a wide range of material properties, for example, NbS$_2$, NbSe$_2$, TaS$_2$, TaSe$_2$, $\beta$-MoTe$_2$, and Td-WTe$_2$ are metals in bulk form,\citep{lebegue_electronic_2009,ding_first_2011,Dawson_Bullett_1987_20_6159__Electronic,augustin_electronic_2000} while
ReS$_2$, ZrS$_2$, MoS$_2$, WS$_2$, MoSe$_2$, WSe$_2$, and $\alpha$-MoTe$_2$ are semiconductors.\citep{lebegue_electronic_2009,ding_first_2011,Zeng_Zhang_2011_50_11093__Single,Tongay_Wu_2014_5_3252__Monolayer,boker_band_2001}
Their layered bulk structure is the stacking of monolayers by weak van der Waals like forces, and the realization of the 2D forms are promised by the in-plane stability of monolayers provided by the strong covalent bonds.
Among the various TMDs, the group-VIB ones (M$=$Mo, W; X$=$S, Se) have been most extensively studied in 2D forms, where both the monolayers and few-layers are proved to be stable in air under room temperature\citep{Geim_Grigorieva_2013_499_419__Van}. Each monolayer is an $\text{X-M-X}$ covalently bonded hexagonal quasi-2D lattice (cf. Fig. \ref{fig:struct}), similar to graphene. Atomically
thin group-VIB TMDs (monolayers, bilayers, etc.) have been prepared in a number
of ways, including mechanical exfoliation from bulk crystals,\citep{novoselov_two_dimensional_2005,splendiani_emerging_2010,mak_atomically_2010,radisavljevic_single_layer_2011}
chemical vapor deposition,\citep{Liu_Li_2012_12_1538__Growth,Zhan_Lou_2012_8_966__Large,Zande_Hone_2013_12_554__Grains,Najmaei_Lou_2013_12_754__Vapour}
and molecular beam epitaxy.\citep{Zhang_Shen_2014_9_111__Direct,Liu_Xie_2014___1407.5744_Dense}
Group-VIB TMDs are attracting great interest as a new class of semiconductors in the 2D limit, having remarkable electronic and optical properties.\citep{Wang_Strano_2012_7_699__Electronics,Xu_Heinz_2014_10_343__Spin}

A number of extraordinary properties associated with the unique quantum
degrees of freedom of electrons have been discovered in 2D group-VIB TMDs.\citep{Xu_Heinz_2014_10_343__Spin}
The conduction and valence bands have degenerate extrema (i.e. valleys)
located at the $K$ and $-K$ points of the hexagonal Brillouin zone
(BZ) (cf. Fig. \ref{fig:struct}c). In particular, monolayers are
found to have a direct band gap at these $\pm K$ valleys,\citep{splendiani_emerging_2010,mak_atomically_2010}
where the optical transitions have a valley dependent selection rule:
interband transitions at $K$ ($-K$) valley couple exclusively to
the right- (left-) circularly polarized light.\citep{Yao_Niu_2008_77_235406__Valley,Xiao_Yao_2012_108_196802__Coupled}
This optical selection rule makes possible quantum control of the
valley pseudospin, including the demonstrated optical generation and
detection of valley polarization\citep{zeng_valley_2012,cao_valley_selective_2012,mak_control_2012}
and valley coherence.\citep{Jones_Xu_2013_8_634__Optical} Moreover,
the valley pseudospin is associated with the valley Hall effect that
makes possible its electric control,\citep{Xiao_Niu_2007_99_236809__Valley,Xiao_Yao_2012_108_196802__Coupled,Mak_McEuen_2014_344_1489__valley}
and the valley magnetic moment for possible magnetic control.\citep{Xiao_Niu_2007_99_236809__Valley,Aivazian_Xu_2014___1407.2645_Magnetic,Srivastava_Imamoglu_2014___1407.2624_Valley,MacNeill_Ralph_2014___1407.0686_Valley}
Similar to spin, these valley phenomena allow the potential use of
valley pseudospin for information processing. In bilayers, electrically
tunable properties of the valley pseudospin and the strong interplay
between the valley pseudospin with the real spin and the layer pseudospin
have also been discovered.\citep{Wu_Xu_2013_9_149__Electrical,Gong_Yao_2013_4_2053__Magnetoelectric,Jones_Xu_2014_10_130__Spin}
2D group-VIB TMDs could provide the platform to explore these various quantum
degrees of freedom of the electrons for future electronic devices
with versatile functionalities.

Because of the direct band gap in the visible frequency range, monolayer group-VIB
TMDs are ideal systems to explore optoelectronic phenomena and applications
in the truly 2D limit. The elementary excitation key to many optoelectronic
phenomena is the exciton, a hydrogen-like bound state formed by a
pair of electron and hole. Under finite doping, this neutral excitation
can bind an extra electron or hole to form a charged exciton, and
such a three-particle bound state is also known as a trion. Recent
experiments show that excitonic physics can be remarkably interesting
in 2D group-VIB TMDs. Photoluminescence (PL) measurements have demonstrated
the continuous evolution from positively charged, to neutral, and
then to negatively charged excitons as a function of doping controlled
by gate.\citep{Mak_Shan_2013_12_207_1210.8226_Tightly,Ross_Xu_2013_4_1474__Electrical,Jones_Xu_2013_8_634__Optical}
First-principles calculations have shown the exotic nature of excitons
in 2D group-VIB TMDs: a strong binding energy of hundreds of meV, and a wave
function largely of the Wannier type (i.e. extended over a large number
of unit cells).\citep{Feng_Li_2012_6_866__Strain,Qiu_Louie_2013_111_216805__Optical}
The calculated binding energy agrees in orders of magnitude with recent
experiments based on optical reflection spectra,\citep{Chernikov_Heinz_2014___1403.4270_Non}
two-photon absorption spectra,\citep{He_Shan_2014___1406.3095_Tightly,Ye_Zhang_2014___1403.5568_Probing,Zhu_Cui_2014___1403.5108_Exciton,Wang_Urbaszek_2014___1404.0056_Non}
and scanning tunneling microscopy and spectroscopy.\citep{Zhang_Shih_2014_14_2443__Direct,Ugeda_Crommie_2014___1404.2331_Observation}
The exceptionally large binding energy and the electrostatic tunability
of excitons in 2D group-VIB TMDs strongly suggest a new paradigm for the study of
exciton physics. Moreover, they imply a particularly strong Coulomb
interaction in 2D group-VIB TMDs due to the reduced dielectric screening in
the 2D geometry,\citep{Qiu_Louie_2013_111_216805__Optical,Ye_Zhang_2014___1403.5568_Probing}
which will lead to interesting many-body phenomena.

Remarkable progresses have been made in devices based upon 2D group-VIB TMDs.
Monolayer MoS$_{2}$ field effect transistors at room temperature have
shown high current on/off ratios, low standby power dissipation and
reasonably good mobility, which significantly promoted the interest
in these 2D materials for device applications.\citep{radisavljevic_single_layer_2011}
Various approaches in fabrication of high quality 2D TMD transistors
have led to the steady progress in improving the mobilities in these
devices.\citep{Fang_Javey_2012_12_3788__High,Liu_Banerjee_2013_13_1983__Role,Baugher_Jarillo-Herrero_2013_13_4212__Intrinsic,Radisavljevic_Kis_2013_12_815__Mobility,Lee_Im_2014___1406.6779_Extremely}
Light-emitting $p$-$n$ junctions have also been created electrostatically
within TMD monolayers by independent gating adjacent regions of the
layer.\citep{Pospischil_Mueller_2014_9_257__Solar,Baugher_Jarillo-Herrero_2014_9_262__Optoelectronic,Ross_Xu_2014_9_268__Electrically,Zhang_Iwasa_2014_344_725__Electrically}
Ambipolar transport in these devices are clearly demonstrated, and
electroluminescence in the forward biased $p$-$n$ junction has been observed.
Interestingly, the electroluminescence from some devices is found
to be circularly polarized, possibly a manifestation of unbalanced
light emissions from the two valleys controlled by electric gating.\citep{Zhang_Iwasa_2014_344_725__Electrically,Yu_Yao_2014___1406.2931_Nonlinear}

Nanostructures in 2D group-VIB TMDs are also being actively explored, including
nano-flakes,\citep{Helveg_Besenbacher_2000_84_951__Atomic,bollinger_one_dimensional_2001}
nano-ribbons,\citep{li_mos_2008,botello_mendez_metallic_2009,ataca_mechanical_2011,kou_tuning_2012,Wang_Iijima_2010_132_13840__Mixed,Erdogan_Seifert_2012_85_33__Transport,Pan_Zhang_2012_22_7280__Edge,Lu_Zeng_2012_14_13035__Strain,Dolui_Sanvito_2012_6_4823__Electric,Yue_Li_2012_24_335501__Bandgap,Sagynbaeva_Ahuja_2014_25_165703__Tweaking,Chu_Zhang_2014_89_155317__Spin}
grains and grain boundaries,\citep{Zande_Hone_2013_12_554__Grains,Najmaei_Lou_2013_12_754__Vapour,Huang_Xu_2014___1406.3122_Lateral}
lateral heterostructures,\citep{Huang_Xu_2014___1406.3122_Lateral}
and quantum dots.\citep{Kormanyos_Burkard_2014_4_11034__Spin,LiuGB-QD-NJP}
In particular, on smooth edges or domain boundaries of TMD monolayers
and bilayers, mid-gap metallic states are found with large density
of states.\citep{Helveg_Besenbacher_2000_84_951__Atomic,bollinger_one_dimensional_2001,Zande_Hone_2013_12_554__Grains,Liu_Xie_2014___1407.5744_Dense}
These metallic states are catalytically active for hydrodesulfurization
and hydrogen evolution reaction, and nano flakes of 2D group-VIB TMDs are of
high interest for these catalytic applications. \citep{Zhuang_Hennig_2013_117_20440__Computational,Besenbacher_Topsoe_2008_130_86__Recent,Byskov_Topsoe_1999_187_109__DFT,Lauritsen_Besenbacher_2001_197_1__Atomic,Lauritsen_Besenbacher_2004_221_510__Atomic,Lauritsen_Besenbacher_2004_224_94__Hydrodesulfurization,Hinnemann_Norskov_2005_127_5308__Biomimetic,Jaramillo_Chorkendorff_2007_317_100__Identification,Kibsgaard_Jaramillo_2012_11_963__Engineering,Kibsgaard_Besenbacher_2006_128_13950__Cluster,Walton_Besenbacher_2013_308_306__MoS2,Liu_Xie_2014___1407.5744_Dense}

The above are only a few selected examples of the appealing features
of 2D group-VIB TMDs, which are being discovered at a rapid pace. They have
made these 2D materials an ideal platform to explore a variety of applications,
as well as an excellent building block for van der Waals heterostructures
that may lead to even richer physics.\citep{Geim_Grigorieva_2013_499_419__Van}

Underlying the extraordinary properties of 2D group-VIB TMDs are their electronic
structures. While 2D group-VIB TMDs are widely considered as the gapped counterpart
of graphene with the same hexagonal 2D lattice, the electronic structures
are much more complicated. At least a total of eleven atomic orbitals in each unit cell -- five $d$-orbitals from the metal atom and six $p$-orbitals from
the two chalcogen atoms -- are considered relevant
in various contexts. The strong spin-orbit coupling (SOC) inherited
from the $d$-orbitals introduces more complexity as well as interesting
spin-dependence in the electronic and optical properties. Unlike graphene where a single-orbital
tight-binding (TB) model well describes the electronic structure of
the active bands over the entire BZ, theoretical models have been
developed for 2D group-VIB TMDs at different levels with their advantages and
limitations, owning to the complex electronic structures. Momentum
space $\kp$ models aim at describing the low energy electrons and
holes near the band edges in the 2D bulk, and the real space TB models
are indispensable for calculations of mesoscopic transport as well
as the study of nanostructures such as quantum dots and nano-ribbons.
Models involving more atomic orbitals may give better description
of the bands over larger $k$-space regions, while simpler models
that capture the essential physics are advantageous in the theoretical
studies of complex phenomena. In light of the rapidly growing research
activities, a review that gives a comprehensive account of the electronic
structures and the theoretical models in 2D group-VIB TMDs, sorting out the
origin of the observed extraordinary properties, is necessary.

This review will focus mainly on the development in the theoretical
understanding of electronic structures in the 2D bulk of group-VIB TMDs including the
band dispersions, electronic and optical band gaps, the spin-orbit
splitting and spin/pseudospin configurations of the energy bands,
the orbital compositions and symmetries of the wave functions, the
selection rules for the optical inter-band transition, interlayer
hopping, heterojunctions, as well as the Berry phase related properties of Bloch electrons
including the Berry curvature and orbital magnetic moment. We will
also give an overview of the various $\kp$ models and TB models developed
so far in describing these electronic structures. Our discussions will focus on the four most extensively studied TMDs: MoS$_2$, WS$_2$, MoSe$_2$, and WSe$_2$ (hereafter TMD refer to these compounds if not stated otherwise).

\section{Electronic structure}

The electronic structures of 2D TMDs are largely determined by their
crystal structures. Bulk TMDs have been studied decades ago ,\citep{Frindt_Frindt_1963_24_1107__optical,Frindt_Yoffe_1963_273_69__Physical,wilson_1969_advphys_18_193_the_1969,verble_lattice_1970,Harper_Edmodson_1971_44_59__Electronic,Bromley_Yoffe_1972_5_759__band,davey_optical_1972,Edmondson_Edmondson_1972_10_1085__Electronic,Mattheiss_Mattheiss_1973_30_784__Energy,mattheiss_band_1973,bullett_electronic_1978}
which are known to crystallize in three different layered structures,
i.e. 1T, 2H (cf. Fig. \ref{fig:struct}d) and 3R (cf. Fig. \ref{fig:struct}e).
1T is not as stable as the 2H and 3R phases for the four group-VIB
TMDs we are focusing on.\citep{Qian_Li_2014___1406.2749_Quantum}
For the 2H and 3R phases, the monolayer has the identical structure
and the difference lies in the stacking order of the monolayers in
the layered structures. The monolayer as the elementary building block
in fact consists of three atomic planes. The top and bottom planes
are chalcogen atoms in a triangular lattice structure, and the middle
plane is another triangular lattice of metal atoms, in trigonal prismatic
coordination (cf. Fig. \ref{fig:struct}b). The three atomic layers
together form a 2D hexagonal lattice, with the A-sublattice being
the metal atom per site and B-sublattice being the two chalcogen atoms
per site. The 2H phase of the bulk crystal has the hexagonal symmetry,
having two monolayers per repeat unit, where the neighboring monolayers
are 180$^{\circ}$ rotation of each other, with the A site of one
layer right on top of the B site of the other, and vice versa (cf.
Fig. \ref{fig:struct}d). The 3R phase has the rhombohedral symmetry,
having three layers per repeat unit, where the neighboring layers
are translation of each other (cf. Fig. \ref{fig:struct}e). First-principles
calculations show that bilayer in 2H stacking is more stable than
the 3R stacking.\citep{WuShiwei} Thin film TMDs exfoliated from the
natural crystals are mostly 2H stacking in the existing studies, even
though 3R stacking is possible, albeit not frequent.\citep{WuShiwei,Suzuki_Iwasa_2014____Valley}

Apart from MS$_2$ and MSe$_2$, bulk MoTe$_2$ can also exist in 2H phase (known as $\alpha$-MoTe$_2$)  below $\approx$ 815$^\circ$C, while at temperature higher than 900$^\circ$C it exists as metallic $\beta$-MoTe$_2$ with monoclinic structure.\citep{boker_valence_band_1999,boker_band_2001} $\alpha$-MoTe$_2$ monolayer has been experimentally realized.\citep{Coleman_et_2011_331_568__Two} Bulk WTe$_2$ exists stably as metallic Td-WTe$_2$ with octahedral coordination, but not in 2H phase.\citep{Dawson_Bullett_1987_20_6159__Electronic,augustin_electronic_2000} 2D form of 2H-WTe$_2$ might exist on some substrate, but has not been realized. Most properties of the 2D MS$_2$ and MSe$_2$ to be discussed applies to 2D 2H-MoTe$_2$ and 2H-WTe$_2$, except that the two  ditellurides have smaller band gap and larger spin-orbit effect than the corresponding disulfides and diselenides.\citep{Liu_Xiao_2013_88_85433__Three}

When group-VIB TMDs are thinned down to monolayer, a critical change
is the crossover from indirect band gap in bulk to direct band gap
in monolayer form. Density functional theory (DFT) calculations have
first pointed to a direct band gap in monolayer WSe$_{2}$ \citep{Voss_Pollmann_1999_60_14311__Atomic}
and MoS$_{2}$,\citep{li_electronic_2007,lebegue_electronic_2009}
located at the corners of the hexagonal BZ, i.e. the $\pm K$ points.
WS$_{2}$ monolayer was also studied a decade ago by both angle-resolved
photoemission spectroscopy (ARPES) experiment and DFT calculation,
which identified its valence band maximum at the $\pm K$ points
but did not draw a conclusion of direct band gap.\citep{klein_electronic_2000,klein_electronic_2001,albe_density_functional_theory_2002}
The first experimental evidence of the direct band gap is from the
PL measurements, where monolayers exhibit orders of magnitude increase
in the quantum efficiency of luminescence compared to the bulk and
multilayers.\citep{splendiani_emerging_2010,mak_atomically_2010}

\begin{figure}
\begin{centering}
\includegraphics[width=8cm]{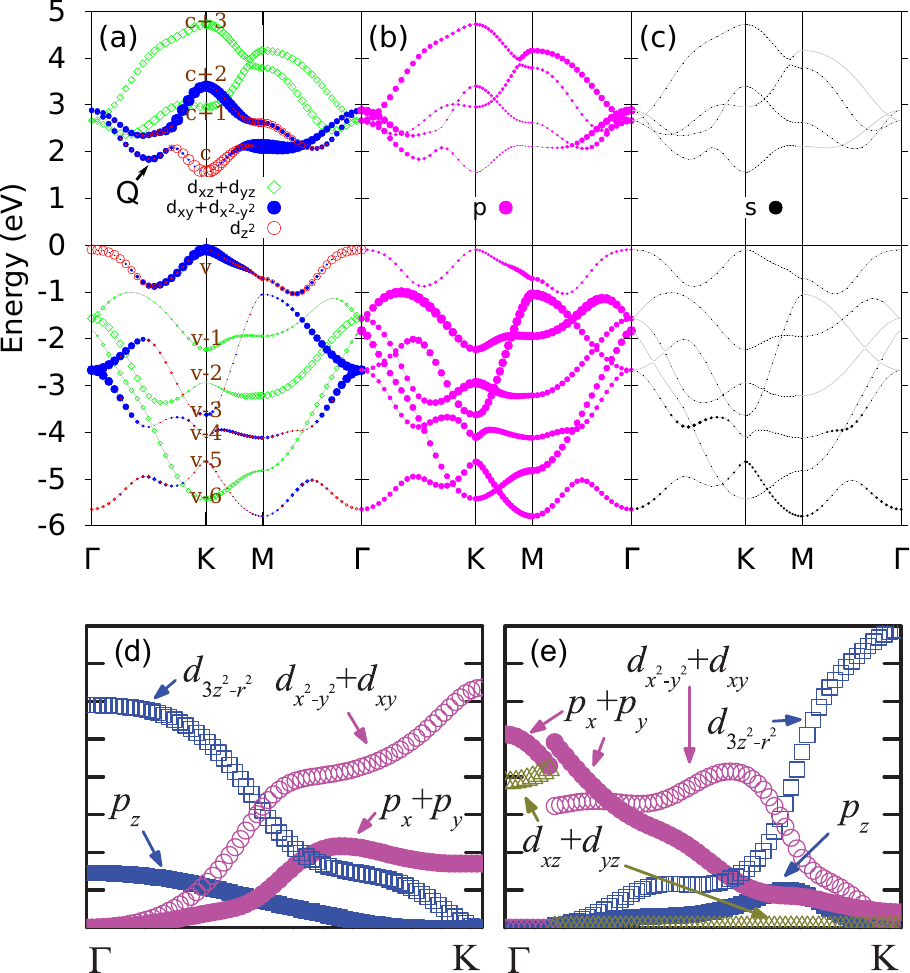}
\par\end{centering}

\caption{(a)-(c) Orbital projected band structures for monolayer MoS$_{2}$
from first-principles calculations without SOC. States at the $K$
points are labelled as from v-6 to c+3. Fermi energy is set to zero.
Symbol size is proportional to its population in corresponding state.
(a) Contributions from Mo-$d$ orbitals: blue dots for $d_{xy}$ and
$d_{x^{2}-y^{2}}$, red open circles for $d_{z^{2}}$, and green open
diamonds for $d_{xz}$ and $d_{yz}$. (b) Total $p$ orbitals, dominated
by S atoms. (c) Total $s$ orbitals. (d)-(e) Orbitally resolved contributions
of Se and W as a function of $k$ along the line $\Gamma$--$K$ for
monolayer WSe$_{2}$: (d) top valence band; (e) lowest conduction
band. (in arbitrary units). (a)-(c) Adapted with permission
from ref. \citenum{Liu_Xiao_2013_88_85433__Three}. Copyright
2013, American Physical Society. (d)-(e) Reproduced with permission from ref. \citenum{zhu_giant_2011}.
Copyright 2011, American Physical Society.\label{fig:orbital}}
\end{figure}

Another critical difference between bulk crystals and monolayer is
the symmetry. The 2H bulk is inversion symmetric, as shown in Fig.
\ref{fig:struct}d. However, when thinned down to monolayer, the unit
cell becomes half of its bulk counterpart, and inversion symmetry
is broken. The crystal symmetry reduces from the space group $D_{6h}^{4}$
for the 2H bulk to the group $D_{3h}^{1}$ for a monolayer. The presence
of inversion symmetry in the bulk and films with even number of layers
and the absence of inversion symmetry in films with odd number of
layers can be best illustrated by a spatial contrast image of second
harmonic generation from a flake with thickness variations.\citep{Zeng_Cui_2013_3_1608__Optical}

In the layered structures, the van der Waals coupling between neighboring
monolayers is much weaker compared to the covalent bonds that form
the monolayer. Below, we will focus first on the monolayers which
are of most current interest because of their direct band gap. Then
we will focus on the bilayers and discuss how the interlayer coupling
can affect the electronic structures. These together will form the
basis to understand the electronic structures of the multilayer thin
films as well as complex van der Waals heterostructures.

\subsection{Monolayers}

\begin{table}
\begin{centering}
\caption{Orbital compositions of Bloch states at the conduction band critical
points $K_{{\rm c}}$ and $Q_{{\rm c}}$, and valence band critical
points $K_{{\rm v}}$ and $\Gamma_{{\rm v}}$ in monolayer TMDs. \label{tab:orb}\medskip{}
}

\par\end{centering}

\centering{}%
\begin{tabular}{ccc}
\hline
state & majority of orbitals & minority of orbitals\tabularnewline
\hline
$K_{{\rm c}}$ & M-$d_{z^{2}}$ & X-$p_{x},p_{y}$\tabularnewline
$K_{{\rm v}}$ & M-$d_{x^{2}-y^{2}},d_{xy}$ &  X-$p_{x},p_{y}$\tabularnewline
$Q_{{\rm c}}$ & M-$d_{x^{2}-y^{2}},d_{xy}$ & M-$d_{z^{2}}$, X-$p_{x},p_{y},p_{z}$\tabularnewline
$\Gamma_{{\rm v}}$ & M-$d_{z^{2}}$ & X-$p_{z}$\tabularnewline
\hline
\end{tabular}
\end{table}

\textbf{Critical points and orbital compositions of energy bands}.
-- Critical points refer to the extrema or saddle points of the energy
bands. They give rise to van Hove singularities in the electronic
density of states, and play crucial roles in determining the transport
and optical phenomena. We consider here the active bands only, i.e.
the top valence band (VB) and the lowest conduction band (CB). In
monolayers, the conduction band minimum (CBM) and the valence band
maximum (VBM) are both located at the corners of the first BZ. The
six corners belong to two inequivalent groups denoted by $K$ and
$-K$ points respectively, where each group has three equivalent corners
related by reciprocal lattice vectors (cf. Fig. \ref{fig:struct}c).
The energetically degenerate but inequivalent band extrema at $K$
and $-K$ constitute a discrete index of carriers, known as the
valley index or pseudospin. Low-energy electrons and holes therefore
have this extra valley degree of freedom in addition to spin. $K$
and $-K$ transform into each other under the time reversal operation.
Hereafter we use $K_{{\rm c}}$ and $K_{{\rm v}}$ to denote the CBM
and VBM at the $\pm K$ points respectively. The calculated effective masses
at $K_{{\rm c}}$ and $K_{{\rm v}}$ are on the order of $\sim 0.5 m_{{\rm 0}}$
($m_{{\rm 0}}$ is the free electron mass) while actual values differ
between calculations depending on the lattice constants and approximation
used.\citep{Ramasubramaniam_Ramasubramaniam_2012_86_115409__Large,Shi_Yakobson_2013_87_155304__Quasiparticle,Chang_Banerjee_2014_115_84506__Ballistic,Wickramaratne_Lake_2014_140_124710__Electronic}

The CB also has six local minima at the low symmetry $Q$ points (cf.
Fig. \ref{fig:struct}c and \ref{fig:orbital}a, also referred as $\Lambda$ points in some literatures\citep{Zhao_Eda_2013_13_5627__Origin,Debbichi_Lebegue_2014_89_205311__Electronic}), while the VB has
a local maximum at the $\Gamma$ point, which are referred to as $Q_{{\rm c}}$
and $\Gamma_{{\rm v}}$ respectively hereafter. These critical points
are also important because they are energetically close to $K_{{\rm c}}$
and $K_{{\rm v}}$ respectively, and in cases such
as under strain or in multilayers, $Q_{{\rm c}}$ may become the
global CBM or $\Gamma_{{\rm v}}$ the global VBM. $K$ and $\Gamma$ are high symmetry
points invariant under the $C_{3}$ operation (the rotation by 2$\pi/3$
around the $z$ axis). With this discrete rotational symmetry, the
dispersions in the neighborhood of these critical points are anisotropic
with the three-fold rotational symmetry (i.e. trigonal warping). Interestingly, the trigonal warping, while not
large, may lead to a new possibility of generating valley and spin
currents that are second order to the electric bias or temperature
gradient. \citep{Yu_Yao_2014___1406.2931_Nonlinear} The six $Q_{{\rm c}}$
valleys can be divided into two groups: $Q$ and $-Q$. The $C_{3}$
operation transforms the three $Q$ (or $-Q$) valleys into each other,
while $Q$ and $-Q$ are related by time reversal operation.

First-principles calculations find that, in monolayers, the 4 bands
above the band gap and the 7 bands below the band gap are predominantly
from the M-$d$ orbitals and X-$p$ orbitals. Other orbitals such
as M-$s,p$ and X-$s$ have negligible contribution. There is perfect
agreement between the first-principles bands and the Wannier bands
constructed using only M-$d$ and X-$p$ orbitals.\citep{feng_intrinsic_2012,Kosmider_Fernandez-Rossier_2013_88_245436__Large}
Orbital compositions of the bulk bands of TMDs are similar to those
of monolayers.\citep{Bromley_Yoffe_1972_5_759__band,mattheiss_band_1973}
Fig. \ref{fig:orbital}a-c shows the orbital compositions of MoS$_{2}$
monolayer and Fig. \ref{fig:orbital}d-e shows those of the WSe$_{2}$
monolayer.

For monolayers, the first-principles calculations also find that the
band-edge states at $K_{{\rm c}}$ and $K_{{\rm v}}$ are predominantly
from the M-$d_{x^{2}-y^{2}},d_{xy},d_{z^{2}}$ orbitals, with some
mixture of X-$p$ orbitals (cf. Tab.\ref{tab:orb}).\citep{Voss_Pollmann_1999_60_14311__Atomic,lebegue_electronic_2009,zhu_giant_2011,Kadantsev_Hawrylak_2012_152_909__Electronic,Chang_Kuo_2013_88_195420__Orbital,Cappelluti_Guinea_2013_88_75409__Tight,Liu_Xiao_2013_88_85433__Three}
For wavefunctions at $Q_{{\rm c}}$ and $\Gamma_{{\rm v}}$, the contributions
from the X-$p_{z}$ orbitals become significant. In fact, the X-$p_{z}$
composition at $Q_{{\rm c}}$ and $\Gamma_{{\rm v}}$ play an important
role in the crossover from the direct to indirect band gap from monolayer
to bulk, because the close distance between X-$p_{z}$ orbitals from neighbouring layers leads to large hopping, which changes the energy of $Q_{{\rm c}}$ and $\Gamma_{{\rm v}}$ substantially.\citep{li_electronic_2007,Cappelluti_Guinea_2013_88_75409__Tight}

Eventhough first-principles calculations have provided valuable insight
into the electronic structures of TMD monolayers, we emphasize that
caution should be applied when interpreting the obtained results,
e.g. the band-edge locations, especially when no experimental data
are available. First-principles calculations have found that the energy
separations between $Q_{{\rm c}}$ and $K_{{\rm c}}$ and between
$\Gamma_{{\rm v}}$ and $K_{{\rm v}}$ depend sensitively on the lattice
constant. It's well known that GGA (LDA) exchange-correlation functional
overestimates (underestimates) lattice constant in the energy minimization
process of first-principles calculations. Hence the lattice constant
determined by GGA (LDA) will introduce artificial tensile (compressive)
strain, which may sometimes give conflicting results. For example,
GGA calculations without SOC for MoS$_{2}$ monolayer give an indirect
gap with $\Gamma_{{\rm v}}$ higher than $K_{{\rm v}}$ by 4 meV, with the GGA lattice constant of 3.19 {\AA}
compared to the bulk experimental value 3.16 \AA.\citep{Liu_Xiao_2013_88_85433__Three,Shi_Yakobson_2013_87_155304__Quasiparticle}
Similarly, LDA gives indirect band gap for MoSe$_{2}$ and WSe$_{2}$
with $Q_{{\rm c}}$ lower than $K_{{\rm c}}$. \citep{Liu_Xiao_2013_88_85433__Three,Horzum_Peeters_2013_87_125415__Phonon,Chang_Kuo_2013_88_195420__Orbital}
In addition to lattice constant, there are other details to affect
the calculated monolayer band structure, such as the GW (quasiparticle
self-energy correction) method,\citep{Hybertsen_Louie_1985_55_1418__First,Louie_Hybertsen_1987_32_31__Theory,Aryasetiawan_Gunnarsson_1998_61_237__GW}
hybrid functional, pseudopotential, and SOC. Turning on SOC, even
in GGA case, will correct MoS$_{2}$ monolayer from indirect to direct
gap. Some GW calculations show that MoS$_{2}$ monolayer has indirect
gap with CBM at $Q_{{\rm c}}$,\citep{Shi_Yakobson_2013_87_155304__Quasiparticle,Huser_Thygesen_2013_88_245309__How}
while others give direct gap.\citep{Qiu_Louie_2013_111_216805__Optical,Liang_Yang_2013_103_42106__Quasiparticle}
The inclusion of the $4s$ (for Mo) and $5s$ (for W) semi-core electrons
is found crucial to obtain the direct gap.\citep{Liang_Yang_2013_103_42106__Quasiparticle}

\begin{table*}[!ht]
\begin{centering}
\caption{Symmetry and orbital compositions of Bloch states $\psi_{n}$ at the
$K$ point in monolayer TMDs. The label $n$
of the states follows from Fig. \ref{fig:orbital}a. Only the dominant
(in bold) and the secondary orbital compositions are shown. $\gamma_{\sigma_{h}}$
is the eigenvalue (or mirror parity) of $\sigma_{h}$. $\gamma_{C_{3}}^{\rho}$
is the eigenvalue of $C_{3}$ (i.e. $C_{3}\psi_{n}=\gamma_{C_{3}}^{\rho}\psi_{n}$)
and $\rho$ is the rotation center located at M, X, or h (cf. Fig.
\ref{fig:struct}a). $\omega\equiv e^{i\frac{2\pi}{3}}.$ The orbital
compositions and $\gamma_{C_{3}}^{\rho}$
all correspond to the states at $K$ in the lattice orientation shown
in Fig. \ref{fig:struct}a, and the $-K$ states are their complex
conjugates. \label{tab:sym}\medskip{}
}

\par\end{centering}

\centering{}%
\begin{tabular}{>{\centering}p{1.5cm}>{\centering}p{1.5cm}>{\centering}p{1.5cm}>{\centering}p{1.5cm}>{\centering}p{1.5cm}>{\centering}p{1.5cm}>{\centering}p{1.5cm}}
\hline
$n$ (in $\psi_{n}$) & Mo-$d$ & S-$p$ & $\gamma_{\sigma_{h}}$ & $\gamma_{C_{3}}^{{\rm M}}$$\vphantom{\Big|}$ & $\gamma_{C_{3}}^{{\rm X}}$ & $\gamma_{C_{3}}^{{\rm h}}$\tabularnewline
\hline
c+3 & $\bm{d_{+1}}$ & $p_{0}$ & $-1$ & $\omega^{*}$ & 1 & $\omega$\tabularnewline
c+2 & $\bm{d_{-2}}$ & $p_{0}$ & $+1$ & $\omega^{*}$ & 1 & $\omega$\tabularnewline
c+1 & $\bm{d_{-1}}$ & $p_{+1}$ & $-1$ & $\omega$ & $\omega^{*}$ & 1\tabularnewline
c & $\bm{d_{0}}$ & $p_{-1}$ & $+1$ & 1 & $\omega$ & $\omega^{*}$\tabularnewline
\hline
v & $\bm{d_{+2}}$ & $p_{+1}$ & $+1$ & $\omega$ & $\omega^{*}$ & 1\tabularnewline
v$-$1 & $d_{+1}$ & $\bm{p_{0}}$ & $-1$ & $\omega^{*}$ & 1 & $\omega$\tabularnewline
v$-$2 & -- & $\bm{p_{-1}}$ & $-1$ & 1 & $\omega$ & $\omega^{*}$\tabularnewline
v$-$3 & $d_{-2}$ & $\bm{p_{0}}$ & $+1$ & $\omega^{*}$ & 1 & $\omega$\tabularnewline
v$-$4 & $d_{+2}$ & $\bm{p_{+1}}$ & $+1$ & $\omega$ & $\omega^{*}$ & 1\tabularnewline
v$-$5 & $d_{0}$ & $\bm{p_{-1}}$ & $+1$ & 1 & $\omega$ & $\omega^{*}$\tabularnewline
v$-$6 & $d_{-1}$ & $\bm{p_{+1}}$ & $-1$ & $\omega$ & $\omega^{*}$ & 1\tabularnewline
\hline
\end{tabular}
\end{table*}

\textbf{Symmetry of band-edge wavefunctions}. -- At the $\pm K$ points,
the wave-vector group is $C_{3h}$ whose generators are $C_{3}$ and
$\sigma_{h}$ (the mirror reflection about the $xy$ plane across
the M atoms). $C_{3h}$ is an Abelian group and has only one-dimensional
irreducible representations. This makes all the Bloch states at $K$
nondegenerate, which have
to be eigenstate of both $C_{3}$ and $\sigma_{h}$. The $\sigma_{h}$
symmetry divides the five M-$d$ orbitals into two sets: the even
set $\{d_{z^{2}},d_{x^{2}-y^{2}},d_{xy}$\} and the odd set $\{d_{xz,}d_{yz}\}$.
X-$p$ orbitals above and below the M-atom plane can also be arranged
in terms of linear superpositions with odd and even $\sigma_{h}$
symmetry respectively. When SOC is not considered, orbitals in the
even set can not couple with those in the odd set. To be eigenstate
of $C_{3}$, orbitals have to combine chirally in spherical harmonic
form, i.e. $d_{\pm2}=\frac{1}{\sqrt{2}}(d_{x^{2}-y^{2}}\pm id_{xy})$,
$d_{\pm1}=\frac{1}{\sqrt{2}}(d_{xz}\pm id_{yz})$, $d_{0}=d_{z^{2}}$,
$p_{\pm1}=p_{x}\pm ip_{y}$, and $p_{0}=p_{z}$. Under rotation about
the atomic center of each orbital, these orbitals transform as
\begin{equation}
C_{3}\alpha_{m}=e^{-i\frac{2m\pi}{3}}\alpha_{m},\ \ (\alpha=p,d;\ m=0,\pm1,\pm2).\label{eq:C3alpha}
\end{equation}
It should be noted that action of $C_{3}$ on the Bloch states includes
rotations of both the constructing atomic orbitals and the plane wave
component in the Bloch functions.\citep{Yao_Niu_2008_77_235406__Valley,cao_valley_selective_2012,LiuGB-QD-NJP}
In Tab. \ref{tab:sym}, we list the main orbital compositions and
the eigenvalues of both $\sigma_{h}$ and $C_{3}$ for all the Bloch
states at the $K$ point. The corresponding eigenvalues
and orbital compositions at the $-K$ point are simply the complex
conjugate since $K$ and $-K$ are time reversal of each other. We
emphasize that the eigenvalue, $\gamma_{C_{3}}^{\rho}$, of $C_{3}$
rotation depends on the position of the rotation center, $\rho$,
which can be M, X, or the h position (the hollow center of the hexagon
formed by M and X, cf. Fig. \ref{fig:struct}a).\citep{LiuGB-QD-NJP}
$\rho=$ M is used in the symmetry analysis of ref. \citenum{Liu_Xiao_2013_88_85433__Three}, while $\rho=$ h
is used in ref. \citenum{cao_valley_selective_2012}
and \citenum{Kormanyos_Falko_2013_88_45416__Monolayer}. The dependence of $C_{3}$ eigenvalues of the Bloch states
on the choice of rotation center can have important consequences,
for example, it gives rise to sensitive dependence of intervalley
coupling strength on the central position of a lateral confinement
potential that also has the $C_{3}$ symmetry.\citep{LiuGB-QD-NJP}

\begin{figure*}[!t]
\begin{centering}
\includegraphics[width=14cm]{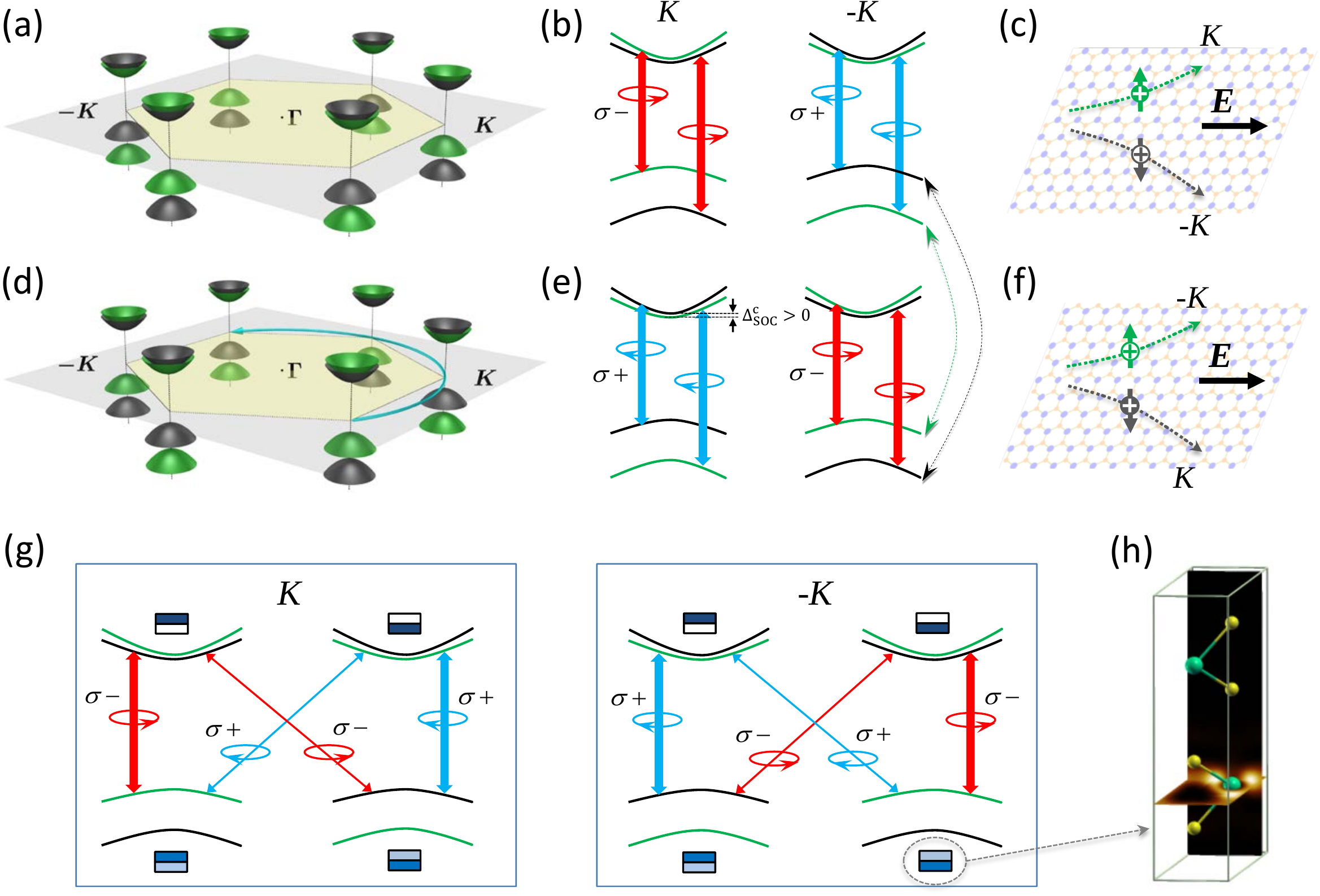}
\par\end{centering}

\caption{Valley and spin physics in TMD monolayers and 2H stacked bilayers.
(a) Schematics of the spin-valley coupled band edges in monolayer WX$_{2}$.
Black (green) color denotes spin-up (-down) bands.
(b) Valley and spin dependent optical
transition selection rules in monolayers.
(c) Valley Hall effect due to the valley contrasting Berry curvature (cf. Fig. \ref{fig:Berry}a-b).
For holes, the valley Hall effect is also a spin Hall effect, as the
spin and valley indices are locked at the VBM. (d)-(f) The same physics
is shown for a WX$_{2}$ monolayer which is a 180$^{\circ}$ rotation
of the one in (a)-(c). The rotation switches the $K$ and $-K$ valleys,
but leaves spin, Berry curvature, and light circular polarization
unchanged. (a)-(f) together illustrates the net effects in two decoupled
monolayers of 2H stacking: (a) and (d) spin splitting with a valley-
and layer- dependent sign; (b) and (e) valley circular dichroism cancels
out in the two layers, while the spin circular dichroism remains; (c)
and (f) valley Hall effect cancels out while the spin Hall effect remains.
(g) Optical transition selection rules in the $K$ and $-K$ valleys
of 2H stacked coupled bilayers. The interlayer hopping (cf. dashed arrows between (b) and (e)) is largely
quenched by the layer dependent spin splitting, so the Bloch states
are predominantly residing an individual layer (illustrated with the rectangular blocks where darker color denotes more occupation).
Thick arrows denote the intralayer optical transitions, and thin arrows
denote the interlayer transitions due to the residue small layer hybridization
of the valence band states. (h) First-principles calculated electron
density map for one valence band Bloch state in a coupled WS$_{2}$
2H bilayer. (b),(e),(g) Adapted with permission from ref. \citenum{Gong_Yao_2013_4_2053__Magnetoelectric}.
(h) Adapted with permission from ref. \citenum{Zeng_Cui_2013_3_1608__Optical}.
Copyright 2013, Nature Publishing Group. \label{fig:bilayer}}
\end{figure*}

\textbf{Valley optical selection rule}. -- One of the most interesting
properties of the TMD monolayers is the valley dependent circularly
polarized selection rules for the optical direct band gap transitions
at the $\pm K$ points.\citep{Yao_Niu_2008_77_235406__Valley,Xiao_Yao_2012_108_196802__Coupled}
This simply follows from the $C_{3}$ eigenvalues of the Bloch states
at the $\pm K$ points. We can establish the following identity for
the interband optical transition matrix element:
\begin{eqnarray}
\bkthree{\psi_{f}}{\hat{P}_{\pm}}{\psi_{i}} & \equiv & \bkthree{\psi_{f}}{C_{3}^{-1}C_{3}\hat{P}_{\pm}C_{3}^{-1}C_{3}}{\psi_{i}}\nonumber \\
 & = & \bkthree{C_{3}\psi_{f}}{C_{3}\hat{P}_{\pm}C_{3}^{-1}}{C_{3}\psi_{i}}\nonumber \\
 & = & e^{i\frac{2\pi}{3}(m_f-m_i \mp1)}\bkthree{\psi_{f}}{\hat{P}_{\pm}}{\psi_{i}}.\label{eq:psiPpsi}
\end{eqnarray}
$\psi_{f}$ and $\psi_{i}$ are the Bloch states at $K$ with
$C_{3}\psi_{f}=e^{-i\frac{2 m_f \pi}{3}}\psi_{f}$, $C_{3}\psi_{i}=e^{-i\frac{2m_i \pi}{3}}\psi_{i}$, where $m_f$ and $m_i$ are integers (c.f. Tab. \ref{tab:sym}).
$\hat{P}$ is the momentum operator, $\hat{P}_{\pm}=\hat{P}_{x}\pm i\hat{P}_{y}$
and $C_{3}\hat{P}_{\pm}C_{3}^{-1}=e^{\mp i\frac{2\pi}{3}}\hat{P}_{\pm}$.
The quantity $|\bkthree{\psi_{f}}{\hat{P}_{\pm}}{\psi_{i}}|^{2}$
characterizes the oscillator strength of the interband transition
with light of $\sigma_{\pm}$ circular polarization. A nonzero $\bkthree{\psi_{f}}{\hat{P}_{\pm}}{\psi_{i}}$
requires $e^{i\frac{2\pi}{3}(m_f -m_i \mp1)}=1$, i.e. ($m_f -m_i \mp1$ modulo
3) $=0$. This is the azimuthal selection rule for allowed interband
optical transitions. We note that the values of $m_f$ and $m_i$ in
Eq. (\ref{eq:psiPpsi}) depend on the choice of rotation center, but
the value ($m_f-m_i $ modulo 3) does not (c.f. Tab. \ref{tab:sym}). The selection rule between
the lowest conduction and top valence bands is of particular interest.
According to Tab. \ref{tab:sym}, the direct band gap optical transition
is allowed at $K$ ($-K$) by the $\sigma_{+}$ ($\sigma_{-}$) polarized
light only (cf. Fig. \ref{fig:bilayer}e). Under time reversal operation,
$K$ is transformed to $-K$, and $\sigma_{-}$ is transformed to
$\sigma_{+}$, so this selection rule does respect the time reversal
symmetry. Inversion operation also transforms $K$ to $-K$, but it
leaves the light circular polarization unchanged. Thus the valley
dependent selection rule conflicts with inversion symmetry, the broken
of this symmetry is the necessary condition for having such selection
rule.\citep{Yao_Niu_2008_77_235406__Valley,Xiao_Yao_2012_108_196802__Coupled}  While the above analysis is applicable to the high symmetry
$\pm K$ points only, first-principles calculations have shown
that the selection rule in fact holds approximately true in a large
neighbourhood of $\pm K$ points.\citep{cao_valley_selective_2012}

The valley optical selection rule makes possible optical pumping and
detection of valley polarization. Circularly polarized light can selectively
inject photocarriers into one of the valleys, while valley polarization
of electron-hole pairs will manifest as circularly polarized luminescence
upon the electron-hole recombination. Initial experimental evidences
of the valley optical selection rule come from the polarization resolved
PL measurement where the light emission is found to have the same
circular polarization as that of the excitation laser.\citep{zeng_valley_2012,cao_valley_selective_2012,mak_control_2012} Another unique consequence of the valley optical selection rule is the possibility to optical generate valley coherence.\citep{Jones_Xu_2013_8_634__Optical} Since linearly polarized photon is the coherent superposition of $\sigma+$ and $\sigma-$ polarized photon, excitation by linearly polarized laser can optically inject electron-hole pair in linear superposition at $K$ and $-K$.

\textbf{Electronic and optical band gaps and excitonic effect}. --
The band gap generally refers to the energy difference between the
CBM and VBM, which can be determined either by transport or optical
measurements. However, the size of the band gap determined from the
two types of measurement differs because of the excitonic effect that
is present in the optical process. The band gap determined from the
transport measurement is also known as \textit{electronic band gap}.
It characterizes single-particle excitations and is defined as the
sum of the energies needed to separately tunnel an electron and a
hole into the system.\citep{Ugeda_Crommie_2014___1404.2331_Observation}
In optical measurement, the absorption of a photon simultaneously
creates an electron in the CB and a hole in the VB, which will bind
through Coulomb interaction into an exciton. The energy required to
creates an exciton is then referred as the \textit{optical band gap}.
Clearly, the difference between the electronic and optical band gaps
corresponds to the binding energy of the exciton, which reflects the
strength of the Coulomb interaction.

\begin{table}[!htbp]
\caption{Optical band gaps in TMD monolayers determined from PL experiments.\label{tab:PLgap}\medskip{}
}

\centering{}%
\begin{tabular}{ccc}
\hline
 & gap & \tabularnewline
\hline
MoS$_{2}$ & 1.83\citep{splendiani_emerging_2010}, 1.90\citep{mak_atomically_2010},  & \tabularnewline
WS$_{2}$ & $1.95$\citep{Zeng_Cui_2013_3_1608__Optical} & \tabularnewline
MoSe$_{2}$ & 1.66\citep{Ross_Xu_2013_4_1474__Electrical} & \tabularnewline
WSe$_{2}$ & $1.64$\citep{Zeng_Cui_2013_3_1608__Optical} & \tabularnewline
\hline
\end{tabular}
\end{table}

Optical band gaps of monolayer TMDs have been determined from PL measurements,
as listed in Tab. \ref{tab:PLgap}. For all four group-VIB TMDs
the band gaps fall into the visible frequency range, making them ideal
systems to explore semiconductor optics and optoelectronic applications.
The substrate and dielectric environment can change the value of the
optical band gap by a few percent.\citep{mak_atomically_2010,splendiani_emerging_2010,Plechinger_Korn_2012_6_126__Low,Bussmann_Schleberger_2012_1474___Electronic,Scheuschner_Maultzsch_2012_249_2644__Resonant,Conley_Bolotin_2013_13_3626__Bandgap,Buscema_CastellanosGomez_2013___1311.3869_effect,Scheuschner_Maultzsch_2014_89_125406__Photoluminescence}
In high quality MoSe$_{2}$ and WSe$_{2}$ samples, the neutral and
charged excitons are found with narrow PL linewidth of a few meV,
while the separation between the neutral and charged excitons is 20-30
meV.\citep{Ross_Xu_2013_4_1474__Electrical,Jones_Xu_2013_8_634__Optical}
This makes possible a reliable characterization of the temperature
dependence of optical gap, which can be well fitted by the standard
semiconductor band gap dependence.\citep{Ross_Xu_2013_4_1474__Electrical}

Electronic band gap has also been measured using scanning tunneling
spectroscopy (STS) and ARPES. For monolayer MoS$_{2}$ on highly ordered
pyrolytic graphite (HOPG), STS measurements give an electronic band
gap of 2.15 eV or 2.35 eV at 77 K depending on the choice of threshold value of tunneling
current,\citep{Zhang_Shih_2014_14_2443__Direct,Chiu_Li_2014___1406.5137_Determination}
while PL measurement on the same sample gives an optical gap of 1.93
eV at 79 K.
For WSe$_{2}$ and WS$_{2}$ monolayers on HOPG, the STS
measurement shows an electronic band gap of 2.51 eV and 2.59 eV respectively
at 77 K.\citep{Chiu_Li_2014___1406.5137_Determination} For monolayer
MoSe$_{2}$ grown by molecular beam epitaxy (MBE) on bilayer graphene,
the STS-determined electronic band gap is 2.18 eV at 5 K, while PL
shows the optical gap of 1.63 eV at 77 K.\citep{Ugeda_Crommie_2014___1404.2331_Observation}
MoSe$_{2}$ grown by MBE on HOPG substrate shows a similar electronic
band gap of 2.1 eV in the STS measurement.\citep{Liu_Xie_2014___1407.5744_Dense}
In the interpretation of the STS measured gap, a possible complication is that $\Gamma_{{\rm v}}$ and $Q_{{\rm c}}$  have much larger weight in the scanning tunneling spectra than $K_{{\rm v}}$ and $K_{{\rm c}}$,\citep{Ugeda_Crommie_2014___1404.2331_Observation} due to the larger density of states as well as larger tunneling coefficient at $\Gamma_{{\rm v}}$ and $Q_{{\rm c}}$. The latter is due to the fact that the wavefunctions at $\Gamma_{{\rm v}}$ and $Q_{{\rm c}}$ have nonnegligible component from X-$p_z$ orbitals, which are closer to the STM tip. This may obscure the attribution of the band edge.
ARPES measurement on heavily doped MBE grown MoSe$_{2}$ monolayer
has reported a much smaller band gap of $\sim1.58$ eV,\citep{Zhang_Shen_2014_9_111__Direct}
implying a large band gap renormalization due to the screening effect
by the carriers.\citep{Ugeda_Crommie_2014___1404.2331_Observation}

The difference between the measured electronic and optical band gaps
point to a large exciton binding energy of hundreds of meV in monolayer
TMDs. An exciton binding energy in the range of 0.3--0.7 eV has also been
inferred in monolayer WX$_2$ from spectral features in two-photon absorption and reflectance
measurement which are attributed as excitonic excited states.\citep{Ye_Zhang_2014___1403.5568_Probing,Zhu_Cui_2014___1403.5108_Exciton,He_Shan_2014___1406.3095_Tightly,Chernikov_Heinz_2014___1403.4270_Non}
Moreover, the charging energy, i.e. energy difference between the
neutral exciton and charged exciton in the PL, is measured to be in
the range of 20--40 meV,\citep{Ross_Xu_2013_4_1474__Electrical,Jones_Xu_2013_8_634__Optical,Mak_Shan_2013_12_207_1210.8226_Tightly}
which is consistent with the above binding energy for 2D excitons. These all point
to exceptionally strong Coulomb interaction, due to the large effective
masses of both electrons and holes and the reduced screening in the
2D limit.\citep{Qiu_Louie_2013_111_216805__Optical,Ugeda_Crommie_2014___1404.2331_Observation}

There have been remarkable efforts in first-principles calculations
of band gaps and the excitonic effects in monolayer TMDs.\citep{Cheiwchanchamnangij_Lambrecht_2012_85_205302__Quasiparticle,Ramasubramaniam_Ramasubramaniam_2012_86_115409__Large,Feng_Li_2012_6_866__Strain,Komsa_Krasheninnikov_2012_86_241201__Effects,Shi_Yakobson_2013_87_155304__Quasiparticle,Berkelbach_Reichman_2013_88_45318__Theory,Qiu_Louie_2013_111_216805__Optical,Huser_Thygesen_2013_88_245309__How}
DFT calculations underestimate the electronic band gap significantly.
The GW method beyond DFT and based on many-body perturbation theory
is generally used to obtain the electronic band gap. To calculate
the exciton binding energy, Bethe-Salpeter equation (BSE)\citep{Salpeter_Bethe_1951_84_1232__Relativistic,Onida_Rubio_2002_74_601__Electronic}
is solved based on the GW method to obtain the optical absorption
spectra and then the exciton binding energy is obtained by subtracting
the optical transition energy from the electronic band gap. The exciton
binding energy from these calculations agrees in order of magnitude
with the experiments. For monolayer MoSe$_{2}$, the experiments and
first-principles calculations agree quantitatively well on the electronic
band gap and the exciton binding energy.\citep{Ugeda_Crommie_2014___1404.2331_Observation}
However, the obtained exciton binding energy depends on the computational
details. $k$ sampling not dense enough overestimates the exciton
binding energy,\citep{Shi_Yakobson_2013_87_155304__Quasiparticle,Molina-Sanchez_Wirtz_2013_88_45412__Effect,Huser_Thygesen_2013_88_245309__How}
while finite interlayer separation underestimates the exciton binding
energy.\citep{Komsa_Krasheninnikov_2012_86_241201__Effects,Huser_Thygesen_2013_88_245309__How}
The underestimation from finite interlayer separation can be solved
by using extrapolation as the interlayer separation going infinite\citep{Komsa_Krasheninnikov_2012_86_241201__Effects}
or by using truncated Coulomb interaction,\citep{Qiu_Louie_2013_111_216805__Optical,Huser_Thygesen_2013_88_245309__How}
and the latter method is considered more reliable.\citep{Huser_Thygesen_2013_88_245309__How}
In addition, including the electron-phonon interaction will result
in red shift of the absorption peak obtained in BSE and affect the
exciton binding energy consequently.\citep{Soklaski_Yang_2014_104_193110__Temperature}
A rather interesting observation from these first-principles calculations
is that the GW correction and the exciton binding energy have comparable
value, giving rise to the accidental agreement between the DFT calculated
gap and the measured optical gap.\citep{Shi_Yakobson_2013_87_155304__Quasiparticle,Qiu_Louie_2013_111_216805__Optical}

The exciton Bohr radius calculated with the GW-BSE method is on the
order of 1 nm,\citep{Feng_Li_2012_6_866__Strain,Qiu_Louie_2013_111_216805__Optical,Berkelbach_Reichman_2013_88_45318__Theory}
still a few times larger than the lattice constant. So the exciton
in monolayer TMDs is of mixed character: the binding energy is comparable
to typical Frenkel excitons, while the wavefunction of the electron-hole
relative motion is largely of the Wannier type, extending over a large
number of unit cells. Correspondingly, the exciton wavefunction is
well localized in momentum space. A bright exciton that can emit photon
has its electron and hole constituents both localized in the $K$
or $-K$ valleys. These valley excitons well inherit the valley optical selection
rule of the band-to-band transition. Namely, the
$K$ ($-K$) valley exciton can be interconverted with $\sigma_{+}$
($\sigma_{-}$) photon. This selection rule forms the basis of the
recently demonstrated optical generation of excitonic valley pseudospin
polarization and coherence.\citep{mak_control_2012,zeng_valley_2012,cao_valley_selective_2012,Jones_Xu_2013_8_634__Optical}
The strong Coulomb interaction in monolayer TMDs also gives rise to
pronounced exchange interaction between the electron and hole constituents
of the exciton, which strongly couples the valley pseudospin of exciton
to its center of mass motion.\citep{Yu_Yao_2014_5_3876__Dirac}

\textbf{Spin-orbital interaction and spin-valley locking}. -- TMDs
have a strong SOC originated from the $d$ orbitals of the metal atoms.
The form of SOC induced spin splitting of bands in monolayers can be fully
determined by symmetry analysis. The first constraint is from $\sigma_{h}$,
the mirror reflection symmetry about the metal atom plane, which dictates
a Bloch state and its mirror reflection to have identical energy.
The $\sigma_{h}$ mirror reflection of an in-plane spin vector is
its opposite, while the mirror reflection of an out-of-plane spin
vector is itself. Thus spin splitting is allowed in out-of-plane ($z$)
direction only, where the spin expectation value of the Bloch states
is either along the $+z$ or $-z$ direction. Secondly, the time reversal
symmetry dictates the spin splittings at an arbitrary pair of momentum
space points $k$ and $-k$ to have identical magnitude but opposite
sign.\citep{Gong_Yao_2013_4_2053__Magnetoelectric} In the neighborhood
of $K$ and $-K$, the SOC then manifests as an effective coupling
between the spin and valley pseudospin, i.e. with the sign of the
spin splitting conditioned on the valley pseudospin. We note that
inversion symmetry would impose a conflicting constraint from that
of the time reversal symmetry, so inversion symmetry breaking in the
monolayer is the necessary condition for having this spin-valley coupling.

\begin{table*}
\caption{SOC splitting at $K_{{\rm v}}$ and $K_{{\rm c}}$ in TMD monolayers
from first-principles calculations.\label{tab:SOCsplit}\medskip{}
}

\centering{}%
\begin{tabular}{cccccc}
\hline
 & \multicolumn{4}{c}{$\Delta_{{\rm SOC}}^{{\rm v}}$ (eV)} & $\Delta_{{\rm SOC}}^{{\rm c}}$ (eV)\tabularnewline
\cline{2-4} \cline{6-6}
 & GGA & HSE\citep{Ramasubramaniam_Ramasubramaniam_2012_86_115409__Large} & GW\citep{Ramasubramaniam_Ramasubramaniam_2012_86_115409__Large} &  & GGA\citep{Liu_Xiao_2013_88_85433__Three}\tabularnewline
\hline
MoS$_{2}$ & 0.148,\citep{zhu_giant_2011,Liu_Xiao_2013_88_85433__Three} 0.146\citep{Ramasubramaniam_Ramasubramaniam_2012_86_115409__Large} & 0.193 & 0.164 &  & $-$0.003\tabularnewline
WS$_{2}$ & 0.430,\citep{Liu_Xiao_2013_88_85433__Three} 0.426,\citep{zhu_giant_2011},
0.425\citep{Ramasubramaniam_Ramasubramaniam_2012_86_115409__Large} & 0.521 & 0.456 &  & $\phantom{+}$0.029\tabularnewline
MoSe$_{2}$ & 0.184,\citep{Liu_Xiao_2013_88_85433__Three} 0.183\citep{zhu_giant_2011,Ramasubramaniam_Ramasubramaniam_2012_86_115409__Large} & 0.261 & 0.212 &  & $-$0.021\tabularnewline
WSe$_{2}$ & 0.466,\citep{Liu_Xiao_2013_88_85433__Three} 0.456,\citep{zhu_giant_2011}
0.461\citep{Ramasubramaniam_Ramasubramaniam_2012_86_115409__Large} & 0.586 & 0.501 &  & $\phantom{+}$0.036\tabularnewline
\hline
\end{tabular}
\end{table*}

First-principles calculations with GGA find that the VB at $K$ point
has a SOC of $\sim0.15$ eV for the MoX$_{2}$ monolayer and $\sim0.45$
eV for WX$_{2}$ monolayer, while the HSE hybrid functional and GW
calculations give slightly larger values (cf. Tab. \ref{tab:SOCsplit}).\citep{zhu_giant_2011,Xiao_Yao_2012_108_196802__Coupled,Ramasubramaniam_Ramasubramaniam_2012_86_115409__Large}
Experimental evidence for such a giant SOC is reflected in PL measurements
where two peaks, attributed as excitons with the holes from the two
split-off spin subbands respectively, are seen with an energy separation
in agreement with the VB spin splitting at $\pm K$.\citep{mak_atomically_2010,Zeng_Cui_2013_3_1608__Optical}
Lately, the ARPES measurement of MBE grown MoSe$_{2}$ gives a direct
evidence of the VBM SOC splitting of $\sim$0.18 eV,\citep{Zhang_Shen_2014_9_111__Direct}
in agreement with the calculations. Because of this giant SOC, the
VBM in monolayer TMDs has the spin index locked with the valley index, i.e.
valley $K$ ($-K$) has only the spin up (down) holes (cf. Fig. \ref{fig:bilayer}a).
Another important consequence of the SOC is the valley dependent optical
selection rule becomes a spin dependent one as well.\citep{Xiao_Yao_2012_108_196802__Coupled}
For example, $\sigma_{+}$ polarized light still excites only the
$K$ valley, but depending on the light frequency, it can either resonantly
excite spin up or down carriers (cf. Fig. \ref{fig:bilayer}e).

The CB at $\pm K$ also has the spin splitting with the same symmetry-dictated
form as the VB. The magnitude, however, is much smaller, about a few
meV for MoS$_{2}$ and tens of meV for other TMDs from first-principles
calculations (cf. Tab. \ref{tab:SOCsplit}), \citep{Cheiwchanchamnangij_Lambrecht_2012_85_205302__Quasiparticle,Kosmider_Fernandez-Rossier_2013_87_75451__Electronic,Kadantsev_Hawrylak_2012_152_909__Electronic,Kormanyos_Falko_2013_88_45416__Monolayer,Roldan_Guinea_2014___1401.5009_Effect,Cheiwchanchamnangij_Dery_2013_88_155404__Strain,Ochoa_Roldan_2013_87_245421__Spin,Liu_Xiao_2013_88_85433__Three,Kosmider_Fernandez-Rossier_2013_88_245436__Large,Kormanyos_Burkard_2014_4_11034__Spin}
and some experimental evidences of this SOC splitting has been reported
recently.\citep{Rivera_Xu_2014___1403.4985_Observation} The small
CB SOC is due to the fact that the CB Bloch states at $\pm K$ are
predominantly formed by the M-$d_{0}$ orbitals where the intra-atomic
$\bm{L\cdot S}$ coupling as the dominant contribution to SOC vanishes
to the leading order. Interestingly, the first-principles calculation
finds an overall sign difference of the CB splitting between MoX$_{2}$
and WX$_{2}$.\citep{Liu_Xiao_2013_88_85433__Three} The CB
splitting originates from the first order effect from the small X-$p_{\pm1}$
compositions, as well as the second-order coupling mediated by the
remote conduction bands consisting of M-$d_{\pm1}$ orbitals (cf. Tab. \ref{tab:sym}).\citep{Ochoa_Roldan_2013_87_245421__Spin,Kormanyos_Falko_2013_88_45416__Monolayer,Liu_Xiao_2013_88_85433__Three}
The two contributions to SOC have opposite sign and their competition
leads to the sign difference between MoX$_{2}$ and WX$_{2}$.\citep{Liu_Xiao_2013_88_85433__Three,Kosmider_Fernandez-Rossier_2013_88_245436__Large}

For the $Q_{{\rm c}}$ valleys of the CB, sizable spin splittings
are also found in the first-principles calculations.\citep{Yu_Yao_2014___1406.2931_Nonlinear}
The three $Q$ ($-Q$) valleys have the same spin splitting as they
are related by the $C_{3}$ rotation, while the splittings at $Q$
and $-Q$ have opposite sign. In the neighborhood of $\Gamma_{{\rm v}}$,
the spin splitting has a quadratic dependence on the wavevector with
a three-fold pattern for the sign. \citep{Yu_Yao_2014___1406.2931_Nonlinear}

\begin{figure}
\centering{}\includegraphics[width=8cm]{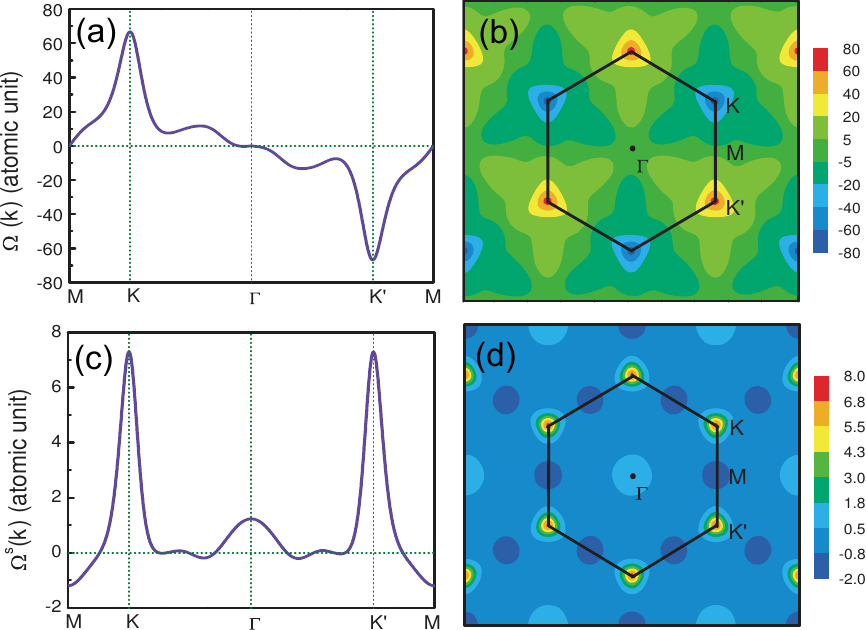}\caption{(a)-(b) First-principles calculated Berry curvature in monolayer MoS$_{2}$.
(a) shows the curvature along the high-symmetry lines, and (b) shows
the distribution in the 2D $k$-plane, in the atomic unit (Bohr$^{2}$).
(c)-(d) First-principles calculated spin Berry curvature in monolayer
MoS$_{2}$. The plots are for the sum of $\Omega_{n}(\vk)$ (or $\Omega_{n}^{s}(\vk)$)
from all valence bands, but the dominant contribution is from the
top valence band of interest. Reproduced with permission from ref.
\citenum{feng_intrinsic_2012}. Copyright 2012, American Physical
Society. \label{fig:Berry}}
\end{figure}

\textbf{Berry phase related properties}. -- The Berry phase effect for a particle lies in the dependence of the internal structure on
the dynamical parameter. In the context of Bloch electrons, it is
the dependence of the periodic part of the Bloch function $u_{n,\vk}(\vr)$
on the wavevector $\vk$.\citep{Xiao_Niu_2010_82_1959__Berry} The Berry curvature and orbital magnetic
moment are two physical quantities that characterize the effect of
Berry phase of electrons in the Bloch bands. The Berry curvature is
defined as: $\bm{\Omega}_{n}(\vk)=i\left\langle \frac{\partial}{\partial\vk}u_{n,\vk}\right|\times\left|\frac{\partial}{\partial\vk}u_{n,\vk}\right\rangle $.
In an applied electric field, the Berry curvature gives rise to an
anomalous velocity $\bm{v}_{\perp}=-\frac{e}{\hbar}\bm{E}\times\bm{\Omega}(\vk)$,
i.e. a Hall effect (cf. Fig. \ref{fig:bilayer}c, f). Thus, Berry
curvature plays the role of a magnetic field in the momentum space.
For 2D crystal, Berry curvature is a pseudo-vector in the out-of-plane
($z$) direction, and its projection along the $z$ axis is:
\[
\Omega_{n}(\vk)=-\sum_{n'(\ne n)}\frac{2{\rm Im}\bkthree{u_{n\vk}}{v_{x}}{u_{n'\vk}}\bkthree{u_{n'\vk}}{v_{y}}{u_{n\vk}}}{(E_{n'}(\vk)-E_{n}(\vk))^{2}},
\]
where $v_{x,y}$ is the velocity operator and $E_{n} (\vk)$ is the energy
dispersion. In monolayer TMDs, sizable Berry curvature is found in
the neighborhood of the $\pm K$ points from theoretical modeling and
first-principles calculations.\citep{Xiao_Yao_2012_108_196802__Coupled,Xu_Heinz_2014_10_343__Spin,cao_valley_selective_2012,feng_intrinsic_2012}
As the $K$ and $-K$ valleys are time reversal of each other, the
Berry curvature must have opposite values at $K$ and $-K$ (cf. Fig.
\ref{fig:Berry}a-b). The Berry curvature is invariant under spatial
inversion that also transform $K$ and $-K$ into each other. Thus,
inversion symmetry breaking is the necessary condition for having
the valley contrasting Berry curvature.

By the valley contrasting Berry curvature, an in-plane electric field
can drive the carriers at the $K$ and $-K$ valleys to the opposite
transverse edges (cf. Fig. \ref{fig:bilayer}c, f). Such a valley
dependent Hall effect is an analog of the spin Hall effect, but with
valley pseudospin playing the role of spin. Valley Hall effect in
monolayer MoS$_{2}$ transistor has been reported recently in experiments.\citep{Mak_McEuen_2014_344_1489__valley}
For hole doped systems, because the spin is locked with the valley
index at the VBM, the valley Hall effect is also accompanied by a
spin Hall effect with the same Hall conductivity (up to a proportionality
constant). For electron
doped case, the spin up and down electrons in the same valley have
slightly different Berry curvature (cf. Tab. \ref{tab:Berry}), which
also gives rise to finite, albeit smaller, spin Hall effect.\citep{Xiao_Yao_2012_108_196802__Coupled,feng_intrinsic_2012}
This is illustrated in Fig. \ref{fig:Berry}c-d, which shows the first-principles calculated spin Berry curvature
\[
\Omega_{n}^{s}(\vk)=-\sum_{n'(\ne n)}\frac{2{\rm Im}\bkthree{\psi_{n\vk}}{j_{x}}{\psi_{n'\vk}}\bkthree{\psi_{n'\vk}}{v_{y}}{\psi_{n\vk}}}{(E_{n'}-E_{n})^{2}},
\]
$j_{x}$ being the spin current operator defined as $\frac{1}{2}(\hat{s}_{z}v_{x}+v_{x}\hat{s}_{z})$.
At the CBM and VBM where spin up and down states are not mixed, the
spin Berry curvature is simply given by $\Omega^{s}(K)=s_{z}\Omega(K)$.
Fig. \ref{fig:Berry}c-d also shows finite spin Berry curvature near
$\Gamma_{{\rm v}}$ where spin Hall effect is also expected.

Berry curvature is in general accompanied by the orbital magnetic
moment, which can be viewed as the self-rotating motion of the electron
wavepacket.\citep{Xiao_Niu_2010_82_1959__Berry} Similar to the spin
magnetic moment, the orbital magnetic moment will lead to Zeeman shift
in a magnetic field. The time reversal symmetry also requires the
orbital magnetic moment to have identical magnitude but opposite signs
at $K$ and $-K$. Therefore, similar to the spin, the two valley
pseudospin states are associated with opposite magnetic moment, making
possible magneto-control of the pseudospin dynamics.\citep{Aivazian_Xu_2014___1407.2645_Magnetic,MacNeill_Ralph_2014___1407.0686_Valley,Srivastava_Imamoglu_2014___1407.2624_Valley}
The valley contrasting Berry curvature and magnetic moment, as well
as the valley and spin dependent optical selection rules, make TMD
monolayers ideal platform for investigating spintronics and valleytronics.

\textbf{Strain effects on band structures}. -- Strain can tune various
physical properties in monolayer TMDs including the band gap, band
edge locations, effective mass, phonon mode, and magnetism, as shown
in theoretical \citep{Lu_Zeng_2012_14_13035__Strain,Yun_Lee_2012_85_33305__Thickness,Li_Li_2012_85_235407__Ideal,Johari_Shenoy_2012_6_5449__Tuning,Yue_Li_2012_376_1166__Mechanical,Scalise_Stesmans_2012_5_43__Strain,Feng_Li_2012_6_866__Strain,Rice_Novoselov_2013_87_81307__Raman,Shi_Yakobson_2013_87_155304__Quasiparticle,Horzum_Peeters_2013_87_125415__Phonon,Chang_Kuo_2013_88_195420__Orbital,Zhang_Schwingenschlogl_2013_88_245447__Giant,Cazalilla_Guinea_2013___1311.6650_Quantum,Qi_Hu_2014_26_256003__Strain,Scalise_Stesmans_2014_56_416__First,Ge_Yao_2014_90_35414__Effect}
and experimental studies. \citep{Conley_Bolotin_2013_13_3626__Bandgap,Rice_Novoselov_2013_87_81307__Raman,Zhu_Urbaszek_2013_88_121301__Strain,He_Shan_2013_13_2931__Experimental,Castellanos_Gomez_Steele_2013_13_5361__Local}
First-principles calculations show that biaxial tensile strain in
MoS$_{2}$ monolayer lifts the energies of $\Gamma_{{\rm v}}$ and
$Q_{{\rm c}}$ points, and under moderate strain ($<2\%$) the band
gap can cross from the direct one ($K_{{\rm v}}$$\leftrightarrow$$K_{{\rm c}}$)
to an indirect one ($\Gamma_{{\rm v}}$$\leftrightarrow$$K_{{\rm c}}$)
\citep{Scalise_Stesmans_2012_5_43__Strain,Li_Li_2012_85_235407__Ideal,Shi_Yakobson_2013_87_155304__Quasiparticle,Yun_Lee_2012_85_33305__Thickness}.
The calculations also find that the band gap decreases with the biaxial
tensile strain and the monolayer may become a metal under larger strain
of 8--10$\%$.\citep{Scalise_Stesmans_2012_5_43__Strain,Li_Li_2012_85_235407__Ideal,Johari_Shenoy_2012_6_5449__Tuning}
In contrast, calculations find that biaxial compressive strain in
MoS$_{2}$ monolayer lowers the energies of $\Gamma_{{\rm v}}$ and
$Q_{{\rm c}}$ points, and the band gap also crosses to an indirect
one ($K_{{\rm v}}$$\leftrightarrow$$Q_{{\rm c}}$, different from
the tensile case). Meanwhile, the band gap increases with compressive strain up to a $\sim2\%$ strain strength
\citep{Yun_Lee_2012_85_33305__Thickness,Lu_Zeng_2012_14_13035__Strain,Chang_Kuo_2013_88_195420__Orbital}
and then decreases with larger strain until becoming metallic.\citep{Scalise_Stesmans_2012_5_43__Strain}
Other TMD monolayers have qualitatively the same behaviors under biaxial
strain.\citep{Johari_Shenoy_2012_6_5449__Tuning,Yun_Lee_2012_85_33305__Thickness,Horzum_Peeters_2013_87_125415__Phonon,Chang_Kuo_2013_88_195420__Orbital}
Uniaxial strain on TMD monolayers are also studied using first-principles
calculations, where similar but quantitatively smaller effects are
found.\citep{Johari_Shenoy_2012_6_5449__Tuning,Chang_Kuo_2013_88_195420__Orbital}
Experiments show that uniaxial tensile strain results in red shift
of both PL and Raman peaks,\citep{He_Shan_2013_13_2931__Experimental,Zhu_Urbaszek_2013_88_121301__Strain,Conley_Bolotin_2013_13_3626__Bandgap}
consistent with first-principles calculations.

Interesting strain effects on valley pseudospin properties in monolayer
TMDs have also been predicted. Uniaxial tensile strain breaks the
$C_{3}$ symmetry and results in an in-plane Zeeman field on the valley
pseudospin of excitons.\citep{Yu_Yao_2014_5_3876__Dirac} A moderate
strain ($\sim1\%$) can lead to a sizable valley pseudospin splitting
of a few meV at zero exciton center-of-mass momentum. It has also
been predicted that inhomogeneous shear strain may lead to valley-contrasting
effective magnetic field in the out-of-plane direction for carriers,
and spin polarized Landau levels can develop.\citep{Guinea_Le_2008_77_205421__Gauge,Levy_Crommie_2010_329_544__Strain,Vozmediano_Guinea_2010_496_109__Gauge,Cazalilla_Guinea_2014_113_77201__Quantum}
These all point to interesting possibilities towards mechanical control
of spin and valley pseudospin of excitons and carriers.

\subsection{Bilayers and heterostructures\label{sub:Bilayers}}

First-principles calculations find that 2H stacking is the most stable
configuration for homostructure bilayer TMDs.\citep{Bhattacharyya_Singh_2012_86_75454_1203.6820v2_Semiconductor,Liu_Lu_2012_116_21556__Tuning,He_Franchini_2014_89_75409__Stacking}
In the 2H stacking, the upper layer is the 180$^{\circ}$ rotation
of the lower one, which is an operation that switches the two valleys
(cf. Fig. \ref{fig:bilayer}a-f). Thus the valley dependent physical
properties of each monolayer, including the valley optical circular
dichroism and valley Hall effect, would average to zero in the pristine 2H bilayer (cf.
Fig. \ref{fig:bilayer}), as required by the restoration of inversion
symmetry (cf. Fig. \ref{fig:struct}d).
Nevertheless, a perpendicular electric field can break the inversion
symmetry, and leads to the emergence of valley circular dichroism
and valley Hall effect.\citep{Yao_Niu_2008_77_235406__Valley,Xiao_Niu_2007_99_236809__Valley,Wu_Xu_2013_9_149__Electrical}
This makes possible a controllable way to tune the valley physical properties by changing the symmetry of the
system. Preliminary evidence
for such tunability has been reported in bilayer MoS$_{2}$ where the PL circular polarization as a signature
of the valley circular dichroism can be switched on and off by the
interlayer bias.\citep{Wu_Xu_2013_9_149__Electrical} The PL circular
polarization has also been used to indicate the inversion symmetry breaking
in the MoS$_{2}$ bilayer.\citep{WuShiwei,Suzuki_Iwasa_2014____Valley} Non-zero
PL circular polarization is generally observed in as-prepared 2H MoS$_{2}$
bilayer in the absence of external gating, implying the existence
of perpendicular electric field from the substrate effect, consistent
with the presence of large n-doping.\citep{Wu_Xu_2013_9_149__Electrical,mak_control_2012}

\begin{table}[!b]
\caption{Conduction and valence band edges of TMD bilayers from first-principles
calculations. \label{tab:bilayer}\medskip{}
}

\centering{}%
\begin{tabular}{ccc}
\hline
bilayer & band edge & refs\tabularnewline
\hline
MoS$_{2}$ & $\Gamma_{{\rm v}}\leftrightarrow Q_{{\rm c}}$ & \citep{splendiani_emerging_2010,Kuc_Heine_2011_83_245213__Influence,Ellis_Scuseria_2011_99_261908__indirect,ramasubramaniam_tunable_2011,Kumar_Ahluwalia_2012_85_186__Electronic,Gong_Yao_2013_4_2053__Magnetoelectric,Debbichi_Lebegue_2014_89_205311__Electronic,Roldan_Guinea_2014___1401.5009_Effect,He_Franchini_2014_89_75409__Stacking}\tabularnewline
 & $\Gamma_{{\rm v}}\leftrightarrow K_{{\rm c}}$ & \citep{Han_Hong_2011_84_45409__Band,Scalise_Stesmans_2012_5_43__Strain,ramasubramaniam_tunable_2011,Yun_Lee_2012_85_33305__Thickness,Liu_Lu_2012_116_21556__Tuning,Bhattacharyya_Singh_2012_86_75454_1203.6820v2_Semiconductor,Lu_Zeng_2012_14_13035__Strain,Cheiwchanchamnangij_Lambrecht_2012_85_205302__Quasiparticle,Zhao_Eda_2013_13_5627__Origin,Zahid_Guo_2013_3_52111__generic}\tabularnewline
\hline
WS$_{2}$ & $\Gamma_{{\rm v}}\leftrightarrow Q_{{\rm c}}$ & \citep{ramasubramaniam_tunable_2011,Yun_Lee_2012_85_33305__Thickness,Kumar_Ahluwalia_2012_85_186__Electronic,Zhao_Eda_2013_13_5627__Origin,Zeng_Cui_2013_3_1608__Optical,Debbichi_Lebegue_2014_89_205311__Electronic,Roldan_Guinea_2014___1401.5009_Effect,He_Franchini_2014_89_75409__Stacking,Gong_Yao_2013_4_2053__Magnetoelectric}\tabularnewline
 & $\Gamma_{{\rm v}}\leftrightarrow K_{{\rm c}}$ & \citep{Kuc_Heine_2011_83_245213__Influence}\tabularnewline
 & $K_{{\rm v}}\leftrightarrow Q_{{\rm c}}$ & \citep{Gong_Yao_2013_4_2053__Magnetoelectric}\tabularnewline
\hline
MoSe$_{2}$ & $\Gamma_{{\rm v}}\leftrightarrow Q_{{\rm c}}$ & \citep{ramasubramaniam_tunable_2011,Kumar_Ahluwalia_2012_85_186__Electronic,Molina-Sanchez_Wirtz_2013_88_45412__Effect,Debbichi_Lebegue_2014_89_205311__Electronic,He_Franchini_2014_89_75409__Stacking,Gong_Yao_2013_4_2053__Magnetoelectric}\tabularnewline
 & $K_{{\rm v}}\leftrightarrow K_{{\rm c}}$ & \citep{Yun_Lee_2012_85_33305__Thickness}\tabularnewline
 & $K_{{\rm v}}\leftrightarrow Q_{{\rm c}}$ & \citep{Gong_Yao_2013_4_2053__Magnetoelectric}\tabularnewline
\hline
WSe$_{2}$ & $K_{{\rm v}}\leftrightarrow Q_{{\rm c}}$ & \citep{Voss_Pollmann_1999_60_14311__Atomic,Yun_Lee_2012_85_33305__Thickness,Zeng_Cui_2013_3_1608__Optical,Debbichi_Lebegue_2014_89_205311__Electronic,Gong_Yao_2013_4_2053__Magnetoelectric,He_Franchini_2014_89_75409__Stacking}\tabularnewline
 & $\Gamma_{{\rm v}}\leftrightarrow Q_{{\rm c}}$ & \citep{Kumar_Ahluwalia_2012_85_186__Electronic}\tabularnewline
 & $\Gamma_{{\rm v}}\leftrightarrow K_{{\rm c}}$ & \citep{Zhao_Eda_2013_13_5627__Origin}\tabularnewline
\hline
\end{tabular}
\end{table}

\textbf{Band edges}. -- The first-principles calculated band structures
of TMD bilayers, in particular the band edges and band gaps, depend
on the parameters and approximations used (cf. Tab. \ref{tab:bilayer}).
Lattice constant and interlayer distance are the two geometry parameters
affecting the band edges most. DFT without the van der Waals correction
will overestimate the interlayer distance in the GGA calculations\citep{ramasubramaniam_tunable_2011,Bhattacharyya_Singh_2012_86_75454_1203.6820v2_Semiconductor}
(may also underestimate a little in LDA\citep{Liu_Lu_2012_116_21556__Tuning,Kumar_Ahluwalia_2012_85_186__Electronic}),
resulting in too small interlayer interactions to describe the bilayers
reasonably. First-principles calculations of bilayer MoS$_{2}$ with
geometry parameters using measured bulk values and the van der Waals
relaxation results (PBE-D2\citep{Grimme_Grimme_2006_27_1787__Semiempirical})
lead to different indirect band gaps of $\Gamma_{{\rm v}}\leftrightarrow Q_{{\rm c}}$
and $\Gamma_{{\rm v}}\leftrightarrow K_{{\rm c}}$ respectively \citep{ramasubramaniam_tunable_2011}.
For WS$_{2}$ and MoSe$_{2}$ bilayers, larger interlayer distance,
i.e. the van der Waals relaxed one compared to the bulk one, will
change the VBM from $\Gamma_{{\rm v}}$ to $K_{{\rm v}}$.\citep{Gong_Yao_2013_4_2053__Magnetoelectric}
Other computational details such as the SOC, exchange correlation
functional, pseudopotential, and GW may also give different results,
e.g. including SOC can change the VBM of WSe$_{2}$ bilayer from $\Gamma_{{\rm v}}$
to $K_{{\rm v}}$.\citep{Zeng_Cui_2013_3_1608__Optical,Debbichi_Lebegue_2014_89_205311__Electronic}
The only direct evidences available are the ARPES studies which show
the VBM at $\Gamma_{{\rm v}}$ for MoS$_{2}$ and MoSe$_{2}$ bilayers\citep{Jin_Osgood_2013_111_106801__Direct,Zhang_Shen_2014_9_111__Direct}.
More experimental studies are needed for determining the band edges
and band gaps in bilayer TMDs under various conditions.

The sensitive band-edge dependence on the lattice constant and interlayer
distance points to significant strain effects. MoS$_{2}$ bilayer
under in-plane biaxial strain are found to behave similarly to its
monolayer counterpart: (i) band gap decreases with the increase of
tensile strain; (ii) under compressive strain the gap first increases
and then decreases; (iii) under large enough strain of either tension
or compression, the bilayer can become metallic.\citep{Scalise_Stesmans_2012_5_43__Strain,Lu_Zeng_2012_14_13035__Strain}
Under out-of-plane compressive strain, TMD bilayers can also undergo
semiconductor-to-metal transition.\citep{Bhattacharyya_Singh_2012_86_75454_1203.6820v2_Semiconductor}
In addition to strain, perpendicular electric field can also decrease
the band gap of TMD bilayers near linearly.\citep{ramasubramaniam_tunable_2011,Liu_Lu_2012_116_21556__Tuning,Zibouche_Heine_2014___1406.5012_Transition}

\begin{figure*}[t]
\begin{centering}
\includegraphics[width=14cm]{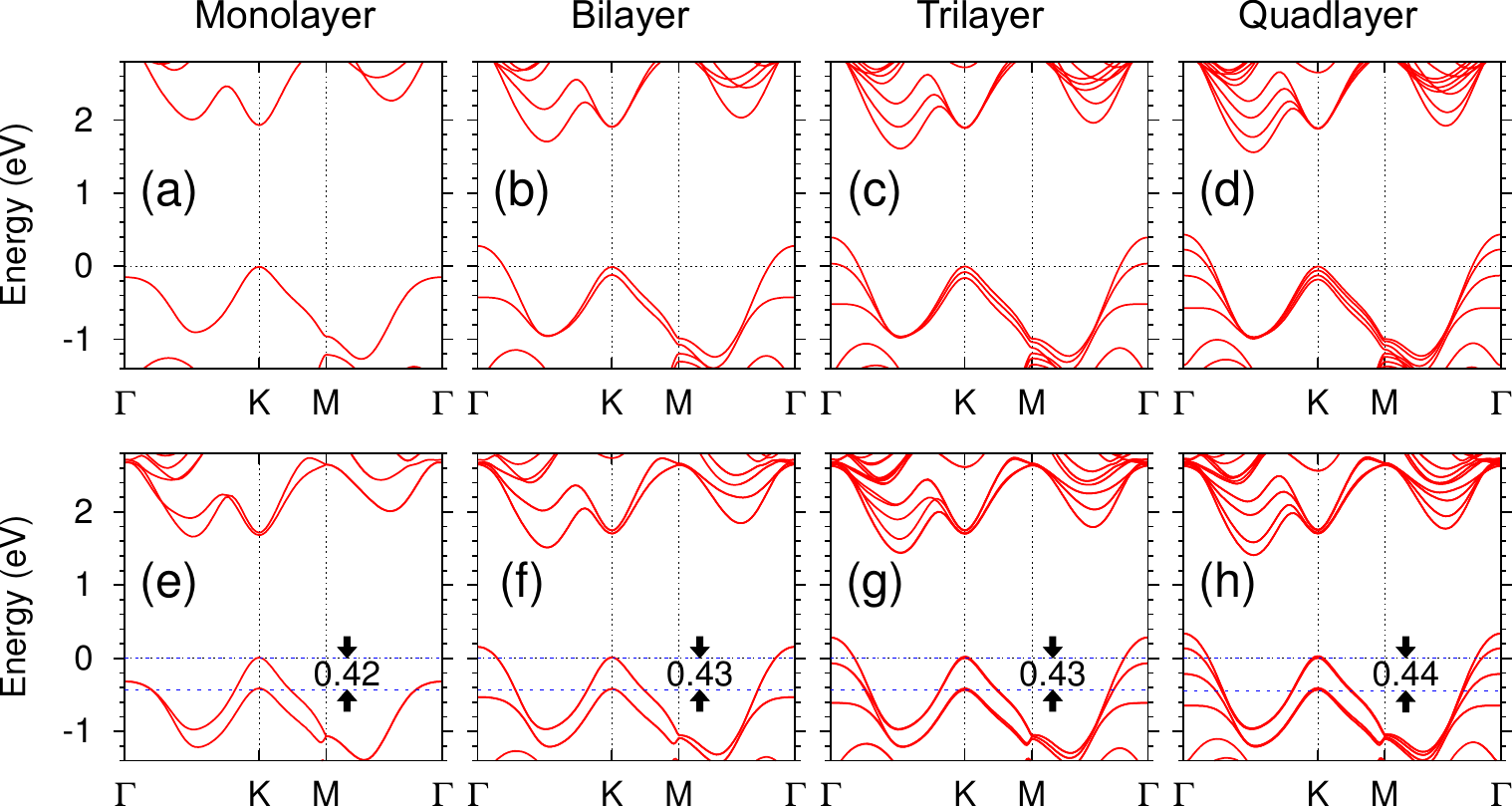}
\par\end{centering}

\caption{First-principles band structures for WS$_{2}$ mono-, bi-, tri-, and
quad-layers without SOC {[}(a)-(d){]} and with SOC {[}(e)-(h){]}.
The valence band splittings at $K$ point are nearly independent of
the number of layers. The experimental bulk lattice constant is used
in the calculations. Adapted with permission from ref. \citenum{Zeng_Cui_2013_3_1608__Optical}.
Copyright 2013, Nature Publishing Group. \label{fig:WS2_1-4layer}}
\end{figure*}

\textbf{Interlayer hopping}. -- The crossover from direct band gap
at monolayer to indirect band gap at bilayer and multilayers is a
consequence of interlayer hopping. It can be seen from Fig. \ref{fig:WS2_1-4layer}a-d
that the band extrema $K_{{\rm v}}$, $\Gamma_{{\rm v}}$ and $Q_{{\rm c}}$
all split with the increase of the number of layers, a direct evidence
of the interlayer hopping. The splitting represents the interlayer
hopping strength. The splitting at $\Gamma_{{\rm v}}$ and $Q_{{\rm {\rm c}}}$
are much larger than that at $K_{{\rm v}}$. This is due to the fact
that the Bloch states at $K_{{\rm v}}$ are predominantly from the
metal $d$ orbitals, while the Bloch states at $\Gamma_{{\rm v}}$
and $Q_{{\rm c}}$ have non-negligible compositions of the chalcogen $p_z$ orbitals
(cf. Fig. \ref{fig:orbital}). The interlayer hopping integral between the $p_z$ orbitals of the nearest neighbor chalcogen atoms is more significant compared to other orbitals. The interlayer hopping at $K_{{\rm v}}$
is then much weaker compared to that at $\Gamma_{{\rm v}}$ and $Q_{{\rm c}}$,
because the two metal planes have larger separation compared to the
two nearest neighbor chalcogen planes from the two layers. With the
increase of the number of layers, the energy of $\Gamma_{{\rm v}}$
is raised and that of $Q_{{\rm c}}$ is lowered significantly, while
the energies of $K_{{\rm c}}$ and $K_{{\rm v}}$ do not change much.
The band gap crosses over to an indirect one at bilayer, and further
decreases with the increase of thickness.

Right at $K_{{\rm c}}$, there is no visible splitting from the interlayer
hopping (Fig. \ref{fig:WS2_1-4layer} a-d). This is dictated by the
rotational symmetry of the 2H bilayer lattice and the Bloch functions
of each monolayer. The interlayer hopping matrix element is $H_{nn'}^{{\rm int}}=\bkthree{\psi_{n}^{{\rm L}}}{\hat{H}^{{\rm int}}}{\psi_{n'}^{{\rm U}}}$,
where $\psi_{n}^{{\rm L}}$ ($\psi_{n'}^{{\rm U}}$) is the Bloch state
at $K$ point in the lower (upper) layer. $\hat{H}^{{\rm int}}$
is the interlayer hopping between the two layers which, in 2H stacking,
is invariant under $C_{3}$ rotation (i.e. $C_{3}\hat{H}^{{\rm int}}C_{3}^{-1}=\hat{H}^{{\rm int}}$).
Thus
\begin{eqnarray}
\!\!\!\! H_{nn'}^{{\rm int}} & \equiv & \bkthree{\psi_{n}^{{\rm L}}}{C_{3}^{-1}C_{3}\hat{H}^{{\rm int}}C_{3}^{-1}C_{3}}{\psi_{n'}^{{\rm U}}}\nonumber \\
 & = & \bkthree{C_{3}\psi_{n}^{{\rm L}}}{C_{3}\hat{H}^{{\rm int}}C_{3}^{-1}}{C_{3}\psi_{n'}^{{\rm U}}}\nonumber \\
 & = & \bkthree{\gamma_{n}^{{\rm L}}\psi_{n}^{{\rm L}}}{\hat{H}^{{\rm int}}}{\gamma_{n'}^{{\rm U}}\psi_{n'}^{{\rm U}}}=(\gamma_{n}^{{\rm L}})^{*}\gamma_{n'}^{{\rm U}}H_{nn'}^{{\rm int}},\label{eq:Hint}
\end{eqnarray}
in which $\gamma_{n}^{{\rm L}}$ and $\gamma_{n'}^{{\rm U}}$ are
the eigenvalues of the $C_{3}$ rotation for $\psi_{n}^{{\rm L}}$
and $\psi_{n'}^{{\rm U}}$ respectively. A nonzero interlayer hopping
matrix element then requires $\gamma_{n}^{{\rm L}}=\gamma_{n'}^{{\rm U}}$. Assuming the lower layer takes the orientation in Fig. \ref{fig:struct}a,
$\gamma_{n}^{{\rm L}}$ is directly given in Tab. \ref{tab:sym},
while $\gamma_{n'}^{{\rm U}}$ is given by the complex conjugates
of those in Tab. \ref{tab:sym} because of the 180$^{\circ}$ rotation
between the two layers that switches $K$ and $-K$. Take the rotation
center to be the M site in the lower layer (X site in the
upper layer), we have $\gamma_{{\rm v}}^{{\rm L}}=\gamma_{{\rm v}}^{{\rm U}}=\omega$
for the VBs, while $\gamma_{{\rm c}}^{{\rm L}}=1$ and $\gamma_{{\rm c}}^{{\rm U}}=\omega^{*}$
for the CBs. These mean that interlayer hopping is allowed between the VBM
states, but is forbidden between the two CBM states at
$K$ point in the lower and upper layers. The strength of the VB interlayer
hopping can be read out from the splitting at $K_{{\rm v}}$ in the
absence of SOC, which is in the order of $\sim100$ meV in all four
TMDs.\citep{Gong_Yao_2013_4_2053__Magnetoelectric} Moreover, for the Bloch states at $K$ point with $n=n'=$c+1,
v-4, or v-6 (cf. Fig. \ref{fig:orbital}), the identical eigenvalues under $C_3$ in the upper and 180$^{\circ}$-rotated lower layers ($\gamma_{n}^{{\rm L}}=\gamma_{n'}^{{\rm U}}=\omega$, cf. Tab. \ref{tab:sym}) imply that these states have finite interlayer hoppings and
hence split in 2H bilayers.

$H_{{\rm cc}}^{{\rm int}}=0$ means that no interlayer hopping exists
to the leading order between the $K_{{\rm c}}$ states in the upper
and lower layers. In fact, in the absence of SOC, $K_{{\rm c}}$ states
in the upper layer and lower layer belong to different irreducible
representation of the $C_{3}$ symmetry, thus interlayer hopping vanishes
to all orders between the $K_{{\rm c}}$ states in the two layers.%
\footnote{We note that small yet finite ($\sim$meV) interlayer splitting exists
for the $K_{{\rm c}}$ states calculated by the VASP software. This
splitting is confirmed to be artificial due to the numerical errors
introduced by VASP. TMD bilayers calculated by a more accurate software,
i.e. WIEN2k, show no splitting for the $K_{{\rm c}}$ state, which
is consistent with the symmetry analysis.%
} We note that this conclusion is unique to the 2H stacking order.
A relative translation or rotation of the two layers will change the
rotational symmetry of $\hat{H}^{{\rm int}}$, allowing a finite interlayer
hopping between the $K_{{\rm c}}$ states.

\textbf{Spin-layer locking effect in $\pm K$ valleys}. -- An interesting
feature in 2H (or 2H-like) stacking is the interlayer hopping at $\pm K$
valleys can be substantially quenched by the SOC. As shown in Fig.
\ref{fig:WS2_1-4layer}e-h for WS$_{2}$, with SOC turned on in the
first-principles calculations, the band dispersions at $K_{{\rm v}}$
become nearly identical for mono-, bi-, tri-, and quad-layers, exhibiting
only a splitting of $\sim0.43$ eV which is nearly the spin-orbit
splitting shown in Fig. \ref{fig:WS2_1-4layer}e. This is due to two
facts: (i) in each monolayer there is an out-of-plane spin splittings
with opposite signs in the $K$ and $-K$ valleys; (ii) the lower
layer is an 180$^{\circ}$ rotation of the upper one in the 2H stacking.
This rotation switches the two valleys in the lower layer, but leave
spin unchanged (cf. Fig. \ref{fig:bilayer}a,d). Thus, the sign of the spin splitting depends on both
the valley index and the layer index.
Since interlayer hopping conserves both the spin and the crystal momentum,
the spin splitting corresponds to an energy cost of the interlayer
hopping (cf. Fig. \ref{fig:bilayer}b,e), which is larger than the
hopping matrix element.\citep{Gong_Yao_2013_4_2053__Magnetoelectric,Jones_Xu_2014_10_130__Spin}
Thus interlayer hopping is effectively quenched, especially for WX$_2$ where SOC is stronger. In 2H bilayers, the Bloch states
in the $\pm K$ valleys are predominantly localized in either the upper
or the lower layer depending on the spin, i.e. the spin index is locked
with the layer index (cf. Fig. \ref{fig:bilayer}g-h). The $ K$
valley physics is essentially that of the two decoupled monolayers.
The valley circular dichroism and valley Hall effect from the two
layers average out, but the spin circular dichroism and the spin Hall
effect from the two layers add constructively (cf. Fig. \ref{fig:bilayer}c,f,g).\citep{Gong_Yao_2013_4_2053__Magnetoelectric}
Circular polarized PL can now come from the spin circular dichroism and is not necessarily an indication of inversion symmetry breaking.\citep{Gong_Yao_2013_4_2053__Magnetoelectric,Jones_Xu_2014_10_130__Spin} Since the layer degree of freedom is associated with the out-of-plane
electric polarization, while spin couples to magnetic field, the spin-layer
locking also leads to a variety of interesting magnetoelectric effects.\citep{Gong_Yao_2013_4_2053__Magnetoelectric,Wu_Xu_2013_9_149__Electrical,Yuan_Iwasa_2013_9_563__Zeeman}

\textbf{Heterostructures of different TMDs}. -- Manually assembled
(stacked) monolayers can lead to a variety of vertical heterostructures
between different group-VIB TMDs \citep{Lee_Kim_2014___1403.3062_Atomically,Furchi_Mueller_2014___1403.2652_Photovoltaic,Cheng_Duan_2014___1403.3447_Electroluminescence,Fang_Javey_2014___1403.3754_Strong,Rivera_Xu_2014___1403.4985_Observation,Chiu_Li_2014___1406.5137_Determination}
and between TMDs and other 2D crystals\citep{Geim_Grigorieva_2013_499_419__Van,Britnell_Novoselov_2013_340_1311__Strong}
, which are of interest because they make possible semiconductor heterojunctions
in the 2D limit.

First-principles calculations have studied various lattice matching
heterostructures of either the AB stacking (M/X site of one layer
right on top of X/M site of the other) or the AA stacking (M/X site
of one layer right on top of M/X site of the other). These heterobilayers
have the same hexagonal BZ as the monolayers. Unlike TMD homobilayers
which all have indirect band gaps, calculations show that some of these lattice matching
heterobilayers exhibit direct band gaps at the $K$ points. For example,
MoS$_{2}$-WSe$_{2}$ and WS$_{2}$-WSe$_{2}$ heterobilayers are
found to be direct band gap semiconductors in the various first-principles
calculations reported.\citep{Terrones_Terrones_2013_3_1549__Novel,Lu_Zeng_2014_6_2879__MoS2,Debbichi_Lebegue_2014_89_205311__Electronic}
MoS$_{2}$-WS$_{2}$ heterobilayer is reported to have indirect band
gap\citep{Terrones_Terrones_2013_3_1549__Novel,Lu_Zeng_2014_6_2879__MoS2,Debbichi_Lebegue_2014_89_205311__Electronic,Komsa_Krasheninnikov_2013_88_85318__Electronic}
or direct band gap \citep{Kosmider_Fernandez-Rossier_2013_87_75451__Electronic}
depending on the interlayer distance adopted in the calculations.
Like the homobilayers, computational details, such as van der Waals
corrections, SOC, GW corrections, matched lattice constant with strain,
etc, may give some different details of the band structures of the
heterobilayers.\citep{Mirhosseini_Felser_2014_89_205301__First,Lu_Zeng_2014_6_2879__MoS2,Komsa_Krasheninnikov_2013_88_85318__Electronic,Kosmider_Fernandez-Rossier_2013_87_75451__Electronic}

An interesting observation from these first-principles calculations
of heterobilayers is that the CB and VB states at the $K$ points
are predominantly localized in an individual layer.
In particular, they exhibit the type-II band-edge alignment with the
CBM and VBM residing in opposite layers.\citep{Debbichi_Lebegue_2014_89_205311__Electronic,Kosmider_Fernandez-Rossier_2013_87_75451__Electronic,Kang_Wang_2013_13_5485__Electronic,Lu_Zeng_2014_6_2879__MoS2,Komsa_Krasheninnikov_2013_88_85318__Electronic}
Such type-II heterojunctions result from the workfunction and band
gap differences between the TMDs. The resultant conduction band offset
(CBO) and valence band offset (VBO) between the monolayers are found
to be a few hundred meV from these calculations. Note that the
CBO and VBO correspond to the energy cost of interlayer hopping, hence
the large CBO and VBO will substantially quench the hybridization
between the layers, making the band structures of the heterobilayers
almost the superposition of that of the constituent monolayers.\citep{Komsa_Krasheninnikov_2013_88_85318__Electronic,Kosmider_Fernandez-Rossier_2013_87_75451__Electronic,Chiu_Li_2014___1406.5137_Determination}

According to the measured bulk lattice constants: 3.160 \AA{}
for MoS$_{2}$,\citep{Coehoorn_Wold_1987_35_6195__Electronic} 3.153
\AA{} for WS$_{2}$,\citep{Schutte_Jellinek_1987_70_207__Crystal}
3.288 \AA{} for MoSe$_{2}$,\citep{Coehoorn_Wold_1987_35_6195__Electronic}
and 3.282 \AA{} for WSe$_{2}$,\citep{Schutte_Jellinek_1987_70_207__Crystal}
lattice matching heterobilayers are  more likely to form between
MoS$_{2}$ and WS$_{2}$, and between MoSe$_{2}$ and WSe$_{2}$.
For other combinations with large mismatch in the lattice constants,
first-principles calculations have found that the interlayer couplings
(van der Waals interactions) are not strong enough to make the two
layers match their lattices.\citep{Kang_Wang_2013_13_5485__Electronic}.
The manually assembled TMD heterostructures in general leads to misalignment
between the layers.

For the manually stacked TMD bilayer heterostructures, optical studies have
shown the spectral features of both the intralayer exciton with electron
and hole from the same layer and the interlayer exciton with electron
and hole from the different layers.\citep{Lee_Kim_2014___1403.3062_Atomically,Furchi_Mueller_2014___1403.2652_Photovoltaic,Cheng_Duan_2014___1403.3447_Electroluminescence,Fang_Javey_2014___1403.3754_Strong,Rivera_Xu_2014___1403.4985_Observation}
For example, in the MoSe$_{2}$-WSe$_{2}$ heterostructures,\citep{Rivera_Xu_2014___1403.4985_Observation}
neutral and charged intralayer excitons from both the MoSe$_{2}$
layer and WSe$_{2}$ layer can be clearly identified in the PL spectral
from the heterojunction region, and they have identical energies to
the ones in isolated MoSe$_{2}$ monolayer and WSe$_{2}$ monolayer.
These further confirm that hybridization between the layers is small for the band edge states.
The interlayer exciton is found to have an energy a few hundred meV
below the intralayer one, consistent with the type-II band edge alignment
and the magnitude of CBO and VBO from the first-principles calculations.

For MoS$_{2}$-WS$_{2}$ and MoS$_{2}$-WSe$_{2}$ heterostructures,
the VBO and CBO have also been directly measured by micro-beam X-ray
photoelectron spectroscopy (XPS) and STS.\citep{Chiu_Li_2014___1406.5137_Determination}
For MoS$_{2}$-WSe$_{2}$ heterostructure, the VBO is directly measured
to be 0.41 eV by the XPS. This VBO can also be inferred from STS measurements
of monolayer MoS$_{2}$ and WSe$_{2}$ both on graphite, where the
deduced VBO is 0.36 eV, consistent with XPS result. And the deduced
value for CBO between MoS$_{2}$-WSe$_{2}$ is 0.52 eV from STS and
0.57 eV combining the VBO from XPS and the band gap measured from
STS. The CBM is found to reside in the MoS$_{2}$ layer and VBM in
the WSe$_{2}$ layer. MoS$_{2}$-WS$_{2}$ heterostructure is shown
to be a similar type-II heterojunction with a smaller VBO of 0.23
eV measured using XPS, and a deduced CBO of $\sim0.4$ eV. WS$_{2}$-WSe$_{2}$
heterostructure is also shown to be a type-II heterojunction with
CBM in the WS$_{2}$ layer and VBM in the WSe$_{2}$ layer, while
both CBO and VBO are $\sim 0.2$ eV.\citep{Chiu_Li_2014___1406.5137_Determination}

In these type-II heterostructures, the observation of interlayer exciton
in the PL spectrum show that there is residue hybridization
between the layers, such that CBM electrons still have a finite albeit
small spatial overlap with the VBM holes residing largely in the opposite
layer. The interlayer exciton is therefore still optically bright,
but has much smaller optical dipole compared to the intralayer exciton.
This gives rise to long recombination lifetime measured for interlayer
exciton which exceeds ns,\citep{Rivera_Xu_2014___1403.4985_Observation}
orders of magnitude longer than the intralayer ones. With electron
and hole constitutions in opposite layers, interlayer exciton correspond
to permanent electric dipole in the out-of-plane direction, evidenced
from the gate dependence of its resonance energy. \citep{Rivera_Xu_2014___1403.4985_Observation}
Repulsive interactions between these dipole-aligned interlayer excitons
are also inferred from the excitation power dependence of the PL spectra.
With the observed ultralong lifetime and the repulsive interaction,\citep{Rivera_Xu_2014___1403.4985_Observation}
the interlayer exciton in TMD heterostructures may provide an ideal
system to explore the exotic phenomenon of excitonic condensation,\citep{Fogler_Novoselov_2014___1404.1418_Indirect}
as well as optoelectronic applications such as the excitonic circuit
and heterostructure lasers. Such interlayer excitons will also form
the basis for optical control of layer degree of freedom in van der
Waals layered structures.

\section{Theoretical models}

Theoretical models have been developed at different levels to describe the complex
electronic structures of TMD monolayers, including both $\kp$ models\citep{Xiao_Yao_2012_108_196802__Coupled,Rostami_Asgari_2013_88_85440__Effective,Kormanyos_Falko_2013_88_45416__Monolayer,Liu_Xiao_2013_88_85433__Three}
and TB models.\citep{Rostami_Asgari_2013_88_85440__Effective,Zahid_Guo_2013_3_52111__generic,Cappelluti_Guinea_2013_88_75409__Tight,Liu_Xiao_2013_88_85433__Three,Roldan_Guinea_2014___1401.5009_Effect}
Among the various models, the simplest and widely used one is
the two-band $\kp$ model describing the neighborhood of the $\pm K$
points \citep{Xiao_Yao_2012_108_196802__Coupled}
\begin{eqnarray}
H_{\kp} & = & at(\tau k_{x}\sigma_{x}+k_{y}\sigma_{y})+\frac{\Delta}{2}\sigma_{z}-\lambda_{{\rm v}}\tau s_{z}\frac{\sigma_{z}-1}{2}\nonumber \\
 & = & \begin{bmatrix}\Delta/2 & at(\tau k_{x}-ik_{y})\\
at(\tau k_{x}+ik_{y}) & -\Delta/2+\lambda_{{\rm v}}\tau s_{z}
\end{bmatrix},
\end{eqnarray}
in which $\sigma_{x/y/z}$ is the Pauli matrix spanning the conduction
and valence states at $\pm K$ points, which are formed respectively
by the $d_{0}$ and $d_{2\tau}$ orbitals respectively. $\tau=\pm1$
is the valley index, and $s_{z}=\pm1$ is the spin index. $a$
is the lattice constant, and $k_{x/y}$ is the wave vector measured
from $\pm K$. The effective hopping integral $t$, the band gap $\Delta$,
and the SOC splitting $2\lambda_{{\rm v}}$ in the VB can all be fitted
from the first-principles band structures in the neighborhood of $K$
points. Interestingly, this two-band $\kp$ model is in fact the massive
Dirac fermion model. This simple model explains why the electron and
hole masses are comparable, since they are largely acquired from the
mutual coupling of the two bands. It also well captures the low-energy
band-edge physics in the $\pm K$ valleys, including the band dispersion,
the giant SOC splitting of the VB, the valley dependent Berry curvature
and orbital magnetic moment, and the valley dependent optical selection rule.\citep{Xiao_Yao_2012_108_196802__Coupled}
Table \ref{tab:Berry} is a comparison of the Berry curvatures from
this model and from first-principles calculations for the four monolayer TMDs,
which shows remarkable agreement.

\begin{table}
\caption{Comparison of the Berry curvatures at VBM and CBM between the first-principles
calculations (the first line) and the $\kp$ model in Ref. \citenum{Xiao_Yao_2012_108_196802__Coupled}
(the second line) for monolayer TMDs. $\Omega_{{\rm v(c)}\uparrow(\downarrow)}$
is the Berry curvature of the valence (conduction) band with spin
$\uparrow$($\downarrow$), given in the unit of Bohr$^{2}$. Reproduced
with permission from ref. \citenum{feng_intrinsic_2012}. Copyright
2013, American Physical Society. \label{tab:Berry}\medskip{}
}

\noindent \centering{}%
\begin{tabular}{ccccc}
\hline
 & MoS$_{2}$ & MoSe$_{2}$ & WS$_{2}$ & WSe$_{2}$\tabularnewline
\hline
$\Omega_{{\rm v\uparrow}}$ & 38.8  &  39.7  &  59.8 &  64.3\tabularnewline
 & 35.3  &  36.5  &  55.4 &  60.0\tabularnewline
$\Omega_{{\rm v\downarrow}}$ & 31.6  &  30.0  &  34.9 &  34.7\tabularnewline
 & 29.5  &  28.4  &  34.2 &  33.3\tabularnewline
$\Omega_{{\rm c\uparrow}}$ & $-35.7$ & $-36.8$ & $-54.7$ & $-59.2$\tabularnewline
 & $-35.3$ & $-36.5$ & $-55.4$ & $-60.0$\tabularnewline
$\Omega_{{\rm c\downarrow}}$ & $-28.8$ & $-27.3$ & $-31.0$ & $-30.8$\tabularnewline
 & $-29.5$ & $-28.4$ & $-34.2$ & $-33.3$\tabularnewline
\hline
\end{tabular}
\end{table}

The SOC in this two-band $\kp$ model is from the intra-atomic contribution
$\bm{L\cdot S}$, thus it vanishes for the CB, missing the
two origins of the small CB splitting: the second-order coupling with
remote M-$d_{\pm1}$ orbitals; and the first-order effect from the
minor X-$p_{\pm1}$ compositions.\citep{Ochoa_Roldan_2013_87_245421__Spin,Kormanyos_Falko_2013_88_45416__Monolayer,Liu_Xiao_2013_88_85433__Three,Kosmider_Fernandez-Rossier_2013_88_245436__Large}
Nevertheless, the CB spin splitting has to take the same symmetry-dictated
form as the VB one as discussed earlier.\citep{Jones_Xu_2014_10_130__Spin} This valley dependent spin
splitting term $\lambda_{{\rm c}}\tau s_{z}$ can be phenomenologically
added to the CB edge for correction.\citep{Rose_Piechon_2013_88_125438__Spin,Kormanyos_Burkard_2014_4_11034__Spin,Ochoa_Roldan_2013_87_245421__Spin}

The two-band $\kp$ model has been widely used to study various properties
of TMD monolayers because of its simplicity.\citep{Li_Niu_2012_110_66803_Unconventional,Lu_Shen_2013_110_16806__Intervalley,Parhizgar_Asgari_2013_87_125401__Indirect,Li_Carbotte_2012_86_205425__Longitudinal,Berkelbach_Reichman_2013_88_45318__Theory,Klinovaja_Loss_2013_88_75404__Spintronics,Li_Carbotte_2013_421_97__Impact,Cai_Niu_2013_88_115140__Magnetic,Rose_Piechon_2013_88_125438__Spin,Shan_Xiao_2013_88_125301__Spin,Zhang_Rana_2014_89_205435__Absorption,Yu_Wu_2014_89_205303__Valley,Cheng_Schwingenschlogl_2014_89_155429__Valley,Sun_Cheng_2014_115_133703__Spin,Rodin_Castro_2013_88_195437__Excitonic,Stroucken_Koch_2014___1404.4238_Evidence}
In the meantime, the simplicity of this model inevitably imposes some
limitations on its applications. For example, it can not account for
the electron-hole asymmetry and the trigonal warping of band dispersion. The two limitations can be
overcome by introducing terms quadratic in $k$ \citep{Rostami_Asgari_2013_88_85440__Effective,Liu_Xiao_2013_88_85433__Three}
or cubic in $k$.\citep{Kormanyos_Falko_2013_88_45416__Monolayer,Liu_Xiao_2013_88_85433__Three}
The corrected models with high-order terms have been used to study
optical conductivity,\citep{Rostami_Asgari_2014_89_115413__Intrinsic}
magneto-optical properties,\citep{Rose_Piechon_2013_88_125438__Spin}
plasmons,\citep{Scholz_Schliemann_2013_88_35135__Plasmons} and spin
relaxation.\citep{Wang_Wu_2014_89_115302__Electron} Apart from the
most studied $\pm K$ valley, the $\Gamma$ valley of VB
can be well described by an effective-mass model nearly unaffected
by SOC.\citep{Kormanyos_Falko_2013_88_45416__Monolayer} For the $Q$
valleys of CB, there lacks simple models because of its low symmetry.

While the $\kp$ approximation aims to describe the electronic structures
only in the neighborhood of a critical point of high symmetry in the
momentum space, TB models can in principle reproduce the band dispersions
in the entire BZ, suitable for studying edge states, finite size systems such as quantum dots, as well as for calculating mesoscopic transport. TB models have been constructed
for TMD monolayers in various approximation levels, with different
number of bands (or the number of orbitals in a unit cell) involved
and different cut-off distances of hoppings. There are (i) 7-band
and 5-band TB models,\citep{Rostami_Asgari_2013_88_85440__Effective}
(ii) 27-band TB model,\citep{Zahid_Guo_2013_3_52111__generic} (iii)
11-band TB model,\citep{Cappelluti_Guinea_2013_88_75409__Tight,Roldan_Guinea_2014___1401.5009_Effect}
and (iv) 3-band TB model.\citep{Liu_Xiao_2013_88_85433__Three} Here
the number of bands is counted in the spinless case, and the actual
number of bands doubles when SOC is added to the model.

(i) The 7-band TB model\citep{Rostami_Asgari_2013_88_85440__Effective}
contains three Mo-$d$ orbitals (Mo-$d_{0},d_{\pm2}$), four S-$p$
orbitals (upper and lower S-$p_{\pm1}$), only Mo-S nearest-neighbor
(NN) hoppings, and overlap integrals. The hopping integrals are in
the Slater-Koster (SK) two-center approximation.\citep{Slater_Koster_1954_94_1498__Simplified}
Then the model is reduced to a 5-band model whose basis states all
have even symmetry under $\sigma_{h}$. The model can describe the
lowest CB and the top VB in the $\pm K$ valleys, while there is large
discrepancy over the rest $k$-space region.\citep{Rostami_Asgari_2013_88_85440__Effective}
Another limitation is that it does not reproduce correctly the order
in energy of other conduction and valence bands. This model has been used to derive the
two-band $\kp$ model with quadratic terms by expansion at $\pm K$
followed by reduction to the two bands in the L\"owdin partitioning
method.\citep{Lowdin_Lowdin_1951_19_1396__Note,Winkler_Winkler_2003____Spin}

(ii) The 27-band TB model \citep{Zahid_Guo_2013_3_52111__generic}
contains the complete set of $sp^{3}d^{5}$ orbitals of the metal
atom and the two chalcogen atoms. The model considers NN SK hoppings
of M-M, M-X, and X-X as well as the corresponding overlap integrals.
The total number of parameters of the model is 96. It well reproduce
all bands from the first-principles bands (in energy range of $-3\sim3$
eV) in the entire BZ . Moreover, this model can apply to monolayer,
bulk, and bilayer using the same set of parameters, all in good agreement
with the first-principles bands. This model is suitable for numerical studies in need of
accurate band descriptions in the entire BZ, but can be too complicated for many studies.

(iii) The 11-band TB model \citep{Cappelluti_Guinea_2013_88_75409__Tight}
contains five M-$d$ orbitals and three $p$ orbitals for each of
the two X atoms. This model also considers NN SK hoppings of M-M,
M-X, and X-X like the 27-band model, while this model uses orthogonal
bases and hence does not need overlap integrals. When SOC is not considered,
the 11-band model can be divided into decoupled $6\times6$ block
and $5\times5$ block which consist of even and odd parity states
respectively under the $\sigma_{h}$ operation. This model is relatively
complete in orbitals for the generally concerned 11 bands (4 above
and 7 below the band gap) and keeps its relative simplicity compared
to the 27-band model. Because both the VB and CB are $\sigma_{h}$-even,
only the parameters for the $\sigma_{h}$-even $6\times6$ block are
fitted.\citep{Cappelluti_Guinea_2013_88_75409__Tight} The fitted
bands agree qualitative well with the first-principles band structures,
but the quantitative discrepancy can be large away from the $K$ and
$\Gamma$ points. With interlayer X-X hoppings considered, this model can be used to qualitatively explain the
direct-to-indirect band gap transition from monolayer to bulk. SOC
interactions have been added to the latest development of this 11-band
model to make it more realistic.\citep{Roldan_Guinea_2014___1401.5009_Effect}

\begin{figure*}[t]
\begin{centering}
\includegraphics[width=12cm]{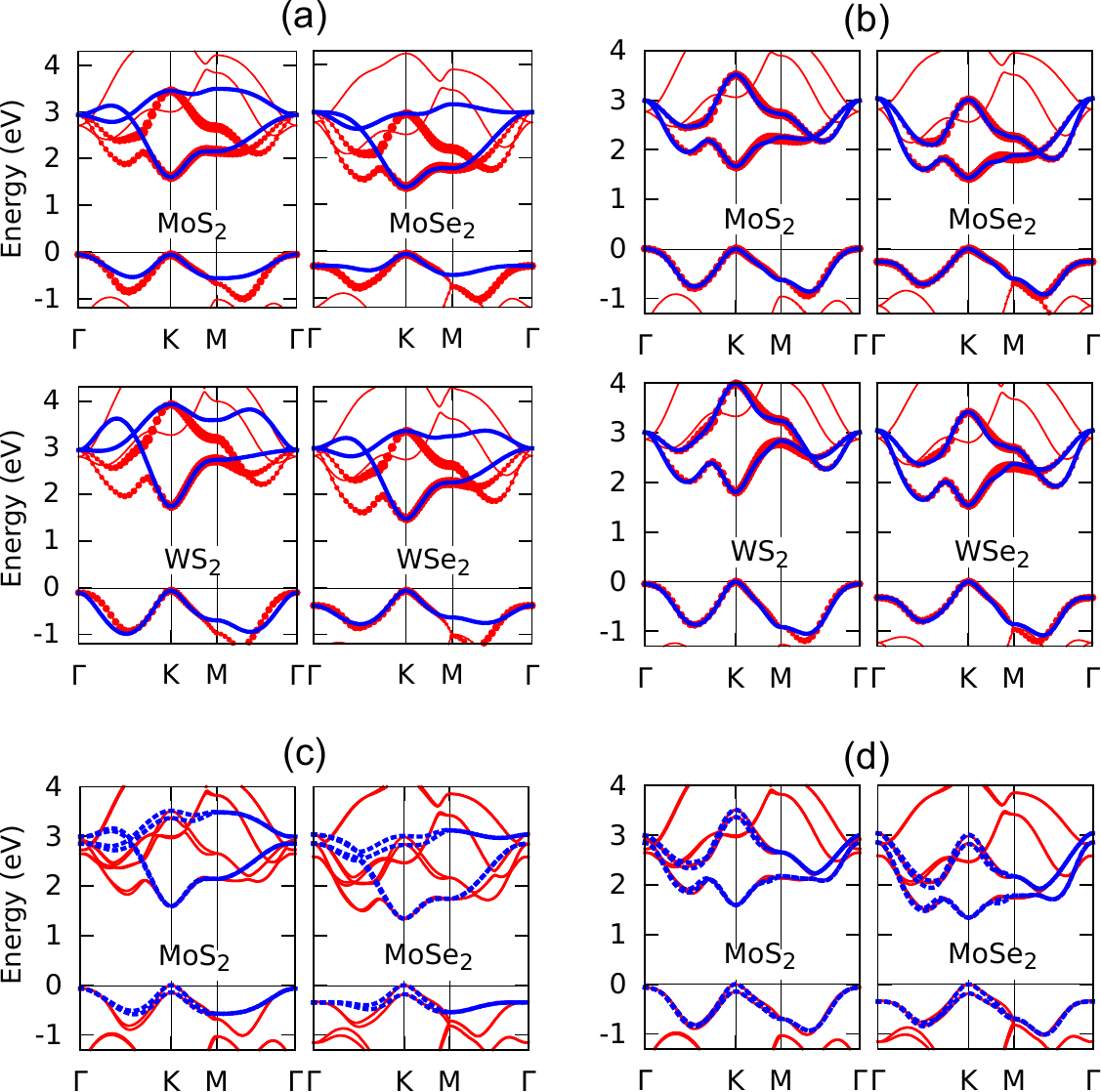}
\par\end{centering}

\caption{Comparison between the bands from the 3-band TB model (blue curves)
and the first-principles ones (red curves and dots) for TMD monolayers.
(a) Bands from NN TB without SOC. (b) Bands from TNN TB without SOC.
Red dots in (a) and (b) show the compositions from the $d_{z^{2}}$,
$d_{xy}$, and $d_{x^{2}-y^{2}}$ orbitals. (c) Bands from NN TB with
SOC. (d) Bands from TNN TB with SOC. Adapted with permission from
ref. \citenum{Liu_Xiao_2013_88_85433__Three}. Copyright 2013, American
Physical Society. \label{fig:fitband}}
\end{figure*}

(iv) The 3-band TB model \citep{Liu_Xiao_2013_88_85433__Three} is
constructed with the three M-$d$ orbitals only, i.e. $d_{z^{2}}$,
$d_{xy}$, and $d_{x^{2}-y^{2}}$. Symmetry-based non-SK M-M hoppings
of both NN interactions and up to the third-nearest-neighbor (TNN)
interactions are considered in this model. Orthogonal bases are used
so that overlap integrals are not needed. Fitting parameters are given
for all MX$_{2}$ monolayers in this model. The bands from the NN
TB model agree well with the first-principles ones only for the VB
and CB in the $\pm K$ valleys (cf. Fig. \ref{fig:fitband}a, c),
while the bands from the TNN TB model agree well with the first-principles
ones in the entire BZ due to the introduction of more hoppings (cf.
Fig. \ref{fig:fitband}b, d). With the absence of X-$p$ orbitals
in the model, physical quantities such as Berry curvature which rely
on the wavefunction structure can be described well only in $k$-space
regions where M-$d$ orbital dominates over the X-$p$ orbitals (e.g.
VB and CB in $\pm K$ valleys) for both the
NN TB and TNN TB models. The SOC splitting of VB can also be well
described. Due to the simplicity of this model, it can
be particularly useful in the study of many-body physics and finite
size systems. This model has been applied to the study
of edge states in MX$_{2}$ nanoribbon,\citep{Chu_Zhang_2014_89_155317__Spin}
quantum dots formed by lateral confinement potential in extended MX$_{2}$
monolayer,\citep{LiuGB-QD-NJP} intercellular orbital magnetic moment,\citep{Srivastava_Imamoglu_2014___1407.2624_Valley}
magnetoelectronic and optical properties,\citep{Ho_Chen_2014_89_155316__Magnetoelectronic}
and magnetoluminescence.\citep{Chu_Zhang_2014_90_45427__Valley}

All the TB models introduced above are capable of describing the low-energy
physics of monolayer TMDs in the $\pm K$ valleys. The 3-band TNN
TB model is the simplest model that gives a reasonably good description
of the top VB and lowest CB in the entire BZ. In order to capture
features such as the CB splitting, M-$d_{xz},d_{yz}$ and X-$p_{x},p_{y}$
orbitals have to be included, as in the 11-band and 27-band models,
which complicates the models inevitably. The TB models in (i)--(iii)
are all based on the SK two-center approximation, while the 3-band
TB model in (iv) is not SK-like but fully based on the symmetry. SK
hoppings omit all contributions from three-center integrals, while
symmetry-based non-SK hoppings include three-center integrals and
make no approximations. This difference can be crucial. For example,
besides 2 onsite energies, the 3-band NN TB model has 6 hoppings.
However, if the 3-band model is constructed in the SK approximation,
there are only 3 hoppings, namely $V_{dd\sigma}$, $V_{dd\pi}$, and
$V_{dd\delta}$. The agreement between the TB bands and first-principles
bands can always be improved by having more fitting parameters, either
by introducing more orbitals or by introducing more hopping integrals.
The success of the 3-band model with the relative simplicity suggests
that going beyond the SK framework can be a viable way of improving
the accuracy of the models.

By introducing interlayer hoppings, all the TB models for monolayer
can be extended to describe bilayer and multilayer homostructures.
Interlayer hoppings are explicitly included in both the 27-band and
the 11-band TB models to describe bilayer or bulk. In general, the
X-$p_{z}$ orbital should be included in a TB model to describe the
interlayer interactions reliably, since it plays a crucial role in
the interlayer hopping over a large range of the BZ, e.g. at $\Gamma_{{\rm v}}$
and $Q_{{\rm c}}$. However, if only the physics in $\pm K$ valleys
are of interest, effective interlayer hoppings can also be included in models constructed with M-$d$ orbitals.

For 2H bilayers, a four-band $\kp$ model has been derived based on
the 3-band TB model of monolayers,\citep{Wu_Xu_2013_9_149__Electrical,Gong_Yao_2013_4_2053__Magnetoelectric}
\begin{equation}
H(\bm{k})=\begin{bmatrix}\Delta & atk_{+}^{\tau} & 0 & 0\\
atk_{-}^{\tau} & -\tau_{z}s_{z}\lambda_{{\rm v}} & 0 & t_{\perp}\\
0 & 0 & \Delta & atk_{-}^{\tau}\\
0 & t_{\perp} & atk_{+}^{\tau} & \tau_{z}s_{z}\lambda_{{\rm v}}
\end{bmatrix}
\end{equation}
in which $k_{\pm}^{\tau}=\tau_{z}k_{x}\pm ik_{y}$. In this model,
only the $k$-independent leading order term is retained for the interlayer
hopping matrix element. The interlayer hopping exists only between
the VB $K_{{\rm v}}$ states in the two layers, while it vanishes
between the CB states at $K_{c}$ due to the symmetry (see discussion
in section \ref{sub:Bilayers}). The magnitude $2t_{\perp}$ can be
extracted from the VB splitting at $K$ from the first-principles
calculations of the bilayer band structures in the absence of SOC.
CB SOC and interlayer bias can be phenomenologically added to this
model. This Hamiltonian has been used to describe the tunable valley
optical circular dichroism and orbital magnetic moment as a function
of interlayer bias,\citep{Wu_Xu_2013_9_149__Electrical} as well as
the magnetoelectric effects in the 2H stacked bilayers.\citep{Gong_Yao_2013_4_2053__Magnetoelectric,Jones_Xu_2014_10_130__Spin}

In the very proximity of $K_{{\rm v}}$, the above bilayer Hamiltonian
simply reduces to:
\begin{equation}
H_{K}=\lambda_{{\rm v}}\tau s_{z}\zeta_{z}+t_{\perp}\zeta_{x},\label{eq:bilayer-K}
\end{equation}
where $\zeta_{z,x}$ is the Pauli matrix for the layer pseudospin:
$\zeta_{z}=1$ for upper layer and $\zeta_{z}=-1$ for lower layer.
$\tau$ and $s_{z}$ are respectively the valley pseudospin and
the real spin. \citep{Gong_Yao_2013_4_2053__Magnetoelectric,Jones_Xu_2014_10_130__Spin}
The SOC term corresponds to the spin splitting in the out-of-plane
direction with a valley and layer dependent sign in 2H bilayer, as
discussed in section \ref{sub:Bilayers}. It manifests as an effective
coupling between the spin, valley pseudospin and the layer pseudospin
in bilayer. $t_{\perp}\zeta_{x}$ is the interlayer hopping term that
conserves the spin and the crystal momentum. The competition of the
SOC term and the interlayer hopping term determines the small hybridization
between the two layers in the proximity of $K_{{\rm v}}$. The energy
eigenstates are associated with a spin- and valley-dependent layer
polarization $\left\langle \zeta_{z}\right\rangle =-\tau s_{z}\frac{\lambda_{{\rm v}}}{\sqrt{\lambda_{{\rm v}}^{2}+t_{\perp}^{2}}}$,
which are found in excellent agreement with the first-principles
wavefunctions.\citep{Gong_Yao_2013_4_2053__Magnetoelectric} The interlayer
hopping and hence layer hybridization at $K_{{\rm c}}$ vanishes due
to symmetry. This difference in the layer hybridization at $K_{{\rm c}}$
and $K_{{\rm v}}$ can result in their different energy shifts in
a perpendicular electric field, which has explained the observed splitting
of interlayer and intralayer trion resonances as a function of interlayer
bias in the PL measurement.\citep{Jones_Xu_2014_10_130__Spin}

\section{Conclusions}

In this review article, we provide an overview of the current understanding
of the various aspects of the electronic structures in 2D group-VIB
TMDs and the theoretical models developed for describing the essential
features. Apart from the two-dimensionality and the visible-frequency-range
direct band gap at the monolayer limit, these 2D semiconductors are
distinguished from all existing systems by their extraordinary properties
including the valley pseudospin of band edge carriers, the valley
dependent Berry phase related properties, the ultra-strong spin-orbit
coupling that lead to the strong interplay between spin and various
pseudospins, and the strong Coulomb interaction evidenced from the
exceptionally large excitonic effects. These imply varieties of interesting
opportunities for the exploration of device applications as well as
fundamental new physics. Although the electronic structures are complex
in general with the involvement of multiple $d$ orbitals of metal
atom and $p$ orbitals of chalcogen atom, the band edge physics in
$\pm K$ valleys which is of most interest for transport and optical
studies is well understood with remarkably simple model that makes
a good example of massive Dirac fermions in the limit of large gap
opening.

We note that this review article has a limited scope, focusing only
on the single particle electronic structures of the 2D bulk. This
is partially due to the lack of understanding of the many-body effects,\citep{Ye_Iwasa_2012_338_1193__Superconducting}
and the electronic structures on the edges, grain boundaries, and
impurities. Even for the 2D bulk, several aspects of the electronic
structures still remain unclear and need further experimental and
theoretical studies to clarify, and we mention a few here. An outstanding
issue of most importance is to determine the difference between the
electronic and optical band gaps, i.e. the exciton binding energy.
First-principles calculations and various experimental studies using
STS, ARPES, optical spectroscopies have agreed on the order of magnitude
of this quantity, while quantitative agreement is yet to be achieved.
The accurate experimental determination of the exciton binding energy
and the exciton excited state are also essential in understanding
the screened Coulomb interaction in these 2D semiconductors.\citep{Qiu_Louie_2013_111_216805__Optical,Ugeda_Crommie_2014___1404.2331_Observation}
The energies of the $Q_{{\rm c}}$ and $\Gamma_{{\rm v}}$ valleys
relative to the CBM and VBM in monolayers remain unclear. These quantities
can be crucial in the study of intervalley relaxation of band edge
carriers which is likely to be mediated by these valleys close by
in energy. The energies of $Q_{{\rm c}}$ and $\Gamma_{{\rm v}}$
from first-principles calculations are various due to the choice of
approximations and the lack of accurate information on lattice constant.
The same problem has also led to the conflicting conclusions on the
indirect band gap and VBM and CBM in bilayers from the first-principles
calculations. ARPES and STS can provide powerful approaches to determine the energies
of these valleys.\citep{Zhang_Shen_2014_9_111__Direct,Ugeda_Crommie_2014___1404.2331_Observation,Zhang_Shih_2014_14_2443__Direct}


A lot of recent experimental efforts have been placed on the study
of heterostructures between different 2D TMDs,\citep{Lee_Kim_2014___1403.3062_Atomically,Furchi_Mueller_2014___1403.2652_Photovoltaic,Cheng_Duan_2014___1403.3447_Electroluminescence,Fang_Javey_2014___1403.3754_Strong,Rivera_Xu_2014___1403.4985_Observation,Chiu_Li_2014___1406.5137_Determination}
which may lead to even richer possibilities for new physics and device
applications. The understanding of the interlayer hopping process
is indispensable in the exploration of the new properties of the heterostructures.
First-principles calculations of AB and AA stacked heterostructures
have suggested that the hybridization between the layers is weak,
due to the large band offsets revealed from STS and XPS,\citep{Chiu_Li_2014___1406.5137_Determination}
and PL measurement.\citep{Lee_Kim_2014___1403.3062_Atomically,Furchi_Mueller_2014___1403.2652_Photovoltaic,Cheng_Duan_2014___1403.3447_Electroluminescence,Fang_Javey_2014___1403.3754_Strong,Rivera_Xu_2014___1403.4985_Observation}
The order of magnitude of the hopping matrix element is of importance
in determining the optical dipole moment of interlayer exciton, and
its dependence on the twisting angle between the layers is of interest.
First-principles calculations can only address very limited cases
of the heterostructures, and efficient theoretical models are needed
with input from the first-principles calculations.

\section*{Acknowledgments}

WY acknowledges support by the Croucher Foundation under the
Croucher Innovation Award, and the RGC (HKU17305914P, HKU9/CRF/13G)
and UGC (AoE/P-04/08) of Hong Kong SAR. GBL acknowledges support by the
NSFC with Grant No. 11304014, the National Basic Research Program
of China 973 Program with Grant No. 2013CB934500 and the Basic Research
Funds of Beijing Institute of Technology with Grant No. 20131842001
and 20121842003. DX acknowledges support by the National Science Foundation, Office of Emerging Frontiers in Research and Innovation (EFRI - 1433496). YY acknowledges support
by the MOST Project of China with Grants Nos. 2014CB920903 and 2011CBA00100,
the NSFC with Grant Nos. 11174337 and 11225418, the Specialized Research
Fund for the Doctoral Program of Higher Education of China with Grants
No. 20121101110046. XX acknowledges support by DoE, BES, Materials Science and Engineering Division (DE-SC0008145), the National Science Foundation (DMR-1150719), and the National Science Foundation, Office of Emerging Frontiers in Research and Innovation (EFRI - 1433496).


\providecommand*{\mcitethebibliography}{\thebibliography}
\csname @ifundefined\endcsname{endmcitethebibliography}
{\let\endmcitethebibliography\endthebibliography}{}

\end{document}